\documentclass[final,3p,11pt,pdflatex]{elsarticle}

\usepackage[utf8x]{inputenc}      
\usepackage[T1]{fontenc}          
\usepackage{lmodern}

\usepackage[obeyspaces]{url}
\usepackage{amsmath,amssymb,mathtools}
\usepackage{bm}
\usepackage{booktabs,tabularx}
\usepackage{graphicx}
\usepackage{xspace}
\usepackage[usenames]{xcolor}
\usepackage{listings}
\usepackage[absolute]{textpos}
\usepackage[many]{tcolorbox}
\usepackage{xparse}
\usepackage[font=small,labelfont=bf,format=plain,margin=0.05\textwidth]{caption}
\usepackage{bbm}
\usepackage{subcaption}
\usepackage{dsfont}
\usepackage[normalem]{ulem}
\usepackage{dirtree}
\usepackage[text,italic]{esdiff}
\usepackage{multirow}
\usepackage{tabularx}
\usepackage{footnote}

\makesavenoteenv{tabular}
\makesavenoteenv{table}

\renewcommand{\appendix}{%
    \setcounter{section}{0}
    \renewcommand*{\thesection}{\Alph{section}}
}

\newcommand{\minima}[1]{{#1}_{\text{min}}}

\definecolor{green}{rgb}{0.13, 0.55, 0.13}
\definecolor{fscolor}{RGB}{44,118,255}

\usepackage[pdftitle={PhaseTracer},
  pdfauthor={Peter Athron, Csaba Bal\'azs,
    Andrew Fowlie, Yang Zhang},
  pdfkeywords={PhaseTracer, phase transitions, electroweak, baryogenesis},
  bookmarks=false, linktocpage, colorlinks=true,
  allcolors=fscolor]{hyperref}
\biboptions{sort&compress}
\bibstyle{elsarticle-num}

\allowdisplaybreaks

\newcommand{\ptgitrepo}{
  \url{https://github.com/PhaseTracer/PhaseTracer}\xspace}
\newcommand{\pthomepage}{\ptgitrepo}

\newcommand{\ep}{\texttt{Effective\-Potential}\@\xspace}
\newcommand{\pt}{\texttt{Phase\-Tracer}\@\xspace}
\newcommand{\bp}{\texttt{Bubble\-Profiler}\@\xspace}
\newcommand{\minuit}{\texttt{MINUIT}\@\xspace}
\newcommand{\cmake}{\texttt{CMake}\@\xspace}
\newcommand{\cmakeminversion}{2.8.12\@\xspace}

\newcommand{\boostminversion}{1.53.0\@\xspace}

\newcommand{\codetab}[1]{\text{\lstinline[]$#1$}}
\newcommand{\code}[1]{\lstinline[breaklines=true,breakatwhitespace=false,postbreak=,prebreak=,breakindent=0pt]{#1}}
\newcommand{\cosmo}{\texttt{Cos\-mo\-Trans\-itions}\@\xspace}
\newcommand{\cosmoversion}{\texttt{Cos\-mo\-Trans\-itions-2.0.3}\@\xspace}
\newcommand{\bsmpt}{\texttt{BSMPT}\@\xspace}
\newcommand{\bsmptversion}{\texttt{BSMPT-1.1.2}\@\xspace}
\newcommand{\fs}{\texttt{Flexible\-SUSY}\@\xspace}
\newcommand{\vevacious}{\texttt{Vevacious}\@\xspace}
\newcommand{\eigenminversion}{3.1.0\@\xspace}
\newcommand{\gev}{\ensuremath{\,\text{GeV}}}

\newcommand{\nloptminversion}{2.4.1\@\xspace}
\newcommand{\make}{\texttt{Make}\@\xspace}

\newcommand{\figref}[1]{\figurename~\ref{#1}}

\newcommand{\secref}[1]{Section~\ref{#1}}

\newcommand{\tabref}[1]{\tablename~\ref{#1}}

\renewcommand{\refeq}[1]{Eq.~\ref{#1}}

\newcommand{\refcite}[1]{Ref.~\cite{#1}}

\providecommand*{\eu}{\ensuremath{e}} 
\DeclareMathOperator{\re}{Re}

\newcommand{\JB}{J_\text{B}}
\newcommand{\JF}{J_\text{F}}
\newcommand{\JBF}{J_\text{B/F}}
\newcommand{\MSbar}{\ensuremath{\overline{\text{MS}}}\xspace}
\newcommand{\DRbar}{\ensuremath{\overline{\text{DR}}}\xspace}

\newcommand{\ztwo}{\ensuremath{\mathbb{Z}_2}\xspace}

\newcommand{\ifrac}[2]{#1/#2}
\newcommand{\idiff}{\renewcommand{\frac}{\ifrac}\diff}

\newcommand{\DThighlight}[1]{\textcolor{red}{#1}}

\NewDocumentEnvironment{OptionTable}{m m O{llXX}}{%
\table[tbh!]
  \tabularx{\textwidth}{#3}%
    \toprule
    Symbol & Default value & Allowed values & Description \\
    \midrule
}{\endtabularx\caption{#1}\label{#2}\endtable}

\makeatletter
\lstnewenvironment{numlstlisting}[1][]{%
  \lstset{%
    #1,
    numbers=left,
    firstnumber=auto,
    numberstyle=\tiny\sffamily}%
  \csname\@lst @SetFirstNumber\endcsname
}{%
  \csname \@lst @SaveFirstNumber\endcsname
}
\makeatother

\newcommand{\gammaEW}{\gamma_\text{EW}}

\begin{document}
\begin{frontmatter}
\vspace*{0.5cm}
\title{\Large\bf PhaseTracer: tracing cosmological phases and calculating transition properties}

\author[Monash,Nanjing]{Peter Athron}
\author[Monash]{Csaba Bal\'azs}
\author[Nanjing]{Andrew Fowlie}
\author[Monash]{Yang Zhang}
\address[Monash]{ARC Centre of Excellence for Particle Physic, School of Physics and Astronomy, Monash University, Melbourne,
  Victoria 3800, Australia}
\address[Nanjing]{Department of Physics and Institute of Theoretical Physics, Nanjing Normal University, Nanjing, Jiangsu 210023, China}

\begin{abstract}
We present a C++ software package called \pt for mapping out cosmological phases, and potential transitions between them, for Standard Model extensions with  any number of scalar fields. \pt traces the minima of effective potential as the temperature changes, and then calculates the critical temperatures, at which the minima are degenerate. \pt is constructed with modularity, flexibility and practicality in mind. It is fast and stable, and can receive potentials provided by other packages such as \fs. \pt can be useful analysing cosmological phase transitions which played an important role in the very early evolution of the Universe.  If they were first order they could generate detectable gravitational waves and/or trigger electroweak baryogenesis to generate the observed matter anti-matter asymmetry of the Universe. The code can be obtained from \pthomepage.
\end{abstract}

\end{frontmatter}

\begin{textblock*}{10em}(\textwidth,1.5cm)
\raggedleft\noindent\footnotesize
CoEPP--MN--20--3 
\end{textblock*}

\clearpage
\newgeometry{top=2.5cm,left=3cm,right=3cm,bottom=4cm,footskip=3em}
\section*{Program Summary}
\noindent
{\em Program title:} \pt\\[0.5em]
{\em Program obtainable from:} \pthomepage\\[0.5em]
{\em Distribution format:} tar.gz\\[0.5em]
{\em Programming language:} C++\\[0.5em]
{\em Computer:} Personal computer\\[0.5em]
{\em Operating system:} Tested on FreeBSD, Linux, Mac OS X \\[0.5em]
{\em External routines:} Boost library, Eigen library, NLopt library, AlgLib library \\[0.5em]
{\em Typical running time:} $0.01$ ($0.05$) seconds for one (two) scalar fields \\[0.5em]
{\em Nature of problem:} Finding and tracing minima of a scalar potential\\[0.5em]
{\em Solution method:} Local optimisation of judicious guesses\\[0.5em]
{\em Restrictions:} Performance inevitably deteriorates for high numbers of scalar fields
\clearpage
\tableofcontents

\newpage
\section{Introduction}
\label{sec:intro}
Since some fundamental symmetries are expected to break in the early Universe the accompanying phase transitions are critical to our understanding of the phenomenon (see e.g.\ \refcite{Mazumdar:2018dfl} for a review).  These transitions occur when the system of scalar fields transits between two distinct minima, also called vacua, of the effective potential.  If the vacua are separated by a barrier the transition is first order, and if the deeper minimum is charged a symmetry breaks spontaneously.

Finite temperature corrections to effective potentials result in important modifications to the free energy typically restoring spontaneously broken symmetries at high temperatures~\cite{Kirzhnits:1972iw, Kirzhnits:1972ut}. Consequently, if a symmetry is broken at zero temperature, a phase transition probably occurred as the Universe cooled.  We therefore expect that there was an electroweak phase transition associated with electroweak symmetry breaking.  There may have been other phase transitions in the early Universe, such as one that breaks symmetries of a Grand Unified Theory (GUT) that embeds the Standard Model (SM) gauge groups into a unified gauge group~\cite{Pati:1974yy, Fritzsch:1974nn, Georgi:1974my, Gursey:1975ki, Georgi:1974sy}.  Alternatively, other gauge groups may have broken at intermediate scales, for example, extra $U(1)$ gauge groups that are fairly generic predictions of string theory \cite{Cvetic:1996mf, Cvetic:1997wu, Cleaver:1998gc, Cleaver:1998sm, Anastasopoulos:2006da, Cvetic:2011iq} and can solve the $\mu$-problem of the Minimal Supersymmetric Standard Model (MSSM) \cite{Suematsu:1994qm, Cvetic:1995rj, King:2005jy}.  These phase transitions played an important role in the evolution of the Universe and it is vital to understand their detailed mechanisms.

First order electroweak phase transitions are particularly interesting as they could help to satisfy Sakharov's third condition for baryogenesis~\cite{Sakharov:1967dj} --- a departure from thermal equilibrium.  Provided there is sufficient CP violation, they could trigger electroweak baryogenesis and explain the observed baryon asymmetry of the Universe  (see e.g.,~\refcite{Cohen:1993nk, Trodden:1998ym, Morrissey:2012db, White:2016nbo} for reviews).  Determining if an electroweak baryogenesis mechanism in a particular extension of the Standard Model can successfully predict the observed baryon asymmetry of the Universe is rather involved, as described by \refcite{Cohen:1993nk, Trodden:1998ym, Morrissey:2012db, White:2016nbo}.  Nonetheless finding the critical temperature of the phase transition is a very important step in this calculation and many studies have focused on this and on the calculation of the order parameter of the first order phase transition (see for example \refcite{Balazs:2007pf, Balazs:2004ae, Balazs:2013cia, Athron:2019teq}).

First order phase transitions also generate gravitational waves via
collisions, sound waves and turbulence from expanding bubbles of a new
phase (see e.g.,~\refcite{Maggiore:1999vm, Weir:2017wfa,
  Alanne:2019bsm, Schmitz:2020syl}). The recent detection of gravitational
waves~\cite{Abbott:2016blz, TheLIGOScientific:2017qsa, Abbott:2016nmj,
  Abbott:2017vtc, Abbott:2017oio} has opened a new window through
which we can directly access physics beyond the Standard Model
\cite{Vaskonen:2016yiu, Alves:2018jsw, Alves:2019igs, Hashino:2016xoj,
  Beniwal:2017eik, Kang:2017mkl, Beniwal:2018hyi, Croon:2019kpe, Dev:2016feu, Dev:2019njv, Bian:2019zpn,Bian:2019kmg}.
Gravitational waves accompanying first order phase transitions could
be observable at current or future gravitational wave detectors, such
as the Einstein Telescope~\cite{Punturo:2010zz}, the Laser
Interferometer Gravitational Wave Observatory
(LIGO)~\cite{TheLIGOScientific:2014jea}, the Virgo
interferometer~\cite{TheVirgo:2014hva}, the Kamioka Gravitational-Wave
Detector(KAGRA)~\cite{Akutsu:2018axf}, the Cosmic
Explorer~\cite{Reitze:2019iox},  the Laser Interferometer
  Space Antenna (LISA)~\cite{amaroseoane2017laser}, the Deci-hertz
  Interferometer Gravitational wave Observatory
  (DECIGO)~\cite{Kawamura:2006up}, the Big Bang Observer
  (BBO)~\cite{Harry:2006fi} or the Taiji program~\cite{Hu:2017mde}.

This would provide new tests of models beyond the SM (BSM), including electroweak baryogenesis, that are complementary to collider physics experiments and measurements of electric dipole moments (EDMs). Even if the energy scale associated with the phase transition is far higher than that which could be probed in any foreseeable collider experiments, gravitational wave detection is still a possibility.  For example, it was shown in \refcite{Croon:2018kqn} that gravitational waves from the breakdown of a Pati-Salam group~\cite{Pati:1974yy} at about $10^5\gev$ can give rise to detectable signatures at the Einstein Telescope.

The gravitational wave spectrum is determined by thermal quantities
such as the nucleation temperature (and rate) of the bubbles, $T_n$,
the latent heat released during the phase transition (relative to the
radiation energy density of the plasma), $\alpha$, the (inverse)
duration of the transition, $\beta$, and the velocity of the expanding
bubble walls, $v_w$. There are significant subtleties involved in
calculating these and a variety of approaches have been taken in the
literature \cite{Apreda:2001us, Leitao:2012tx, Dorsch:2018pat,
  Ellis:2018mja, Ellis:2019oqb}. Nonetheless, finding the first order
phase transitions and their associated critical temperatures is an
important step towards testable gravitational wave predictions.

To help study phase transitions and their associated phenomenology, we present \pt, a code to analyse the possible phases and phase transitions as the Universe cooled for any scalar potential. There are two other public codes with similar goals: \cosmo~\cite{Wainwright:CosmoTransition} and \bsmpt~\cite{Basler:2018cwe}.\footnote{A \code{C++} version of \cosmo is adopted by \vevacious \cite{Camargo-Molina:2013qva} but the latter is mainly set up to test whether or not the observed electroweak vacuum is unstable and should have transitioned to a deeper underlying minimum.}  Our code is most similar to the former, which facilitated over a hundred studies.  Compared to \cosmo, the advantages of \pt are speed, as it is written in \code{C++} rather than \code{Python}; flexibility, as it can be linked to models in \bsmpt and \fs~\cite{Athron:2014yba,Athron:2016fuq,Athron:2017fvs}; robustness, as it includes a test-suite of known analytic and numerical results and correctly handles discrete symmetries; and lastly, active maintenance and development by several authors.  Our code is public and we encourage contributions (such as repository pull requests) from the community.

\bsmpt and \pt have different strategies for finding and tracing the temperature dependence of vacua.  In the cases that we tested, however, \pt was able to identify more transitions than \bsmpt.  The public code \vevacious uses a numerical polyhedral homotopy continuation method, a powerful way to find all the roots of a system of polynomial equations, to identify vacua at zero temperature.  This technique is more robust than the one used presently in \pt. However it requires the potential to be written in a special symbolic format and can only be used directly to find the tree-level minimum at zero temperature.  After this, adjustments to the minima locations from one-loop Coleman-Weinberg corrections are then found using the \minuit algorithms \cite{JAMES1975343}. Minima at finite temperature are also found via \minuit, using each minimum at zero temperature as a starting point and assuming that these two minima belong to the same phase, which is not always true.

With \pt our aim is to improve the calculation of phase transitions on several fronts.  We intend to locate all minima at zero temperature and a high temperature and find the thermal phases of the potential by following the temperature dependence of the minima.
We attempt to reliably distinguish phases during the evolution of the minima with temperature.
We try to properly map the temperature dependence of \textit{all} phases being aware that minima may appear and disappear with the temperature evolution.  Finally, we aim to properly identify the potential transitions between the identified thermal phases including cases when the potential may have discrete symmetries.

The structure of our paper is as follows. In \secref{sec:QuickStart}
we provide quickstart instructions for installing and running \pt.
Then in \secref{sec:PhysicalProblem} we describe the physics problem
and our numerical methods to solve it, as well as presenting the zero-
and finite-temperature corrections that we include in our
calculations. After this we briefly describe the structure of the code
in \secref{sec:structure} before providing detailed instructions and
examples on how to implement new models in \secref{sec:Implementing_Models}. In \secref{sec:examples} we provide an extensive set of example results
and comparisons between \pt and other codes or analytic results.
Finally in \secref{sec:conclusions} we present our conclusions.

\section{Quick start}
\label{sec:QuickStart}

\subsection{Requirements}

Building \pt requires the following:
\begin{itemize}
\item A \code{C++11} compatible compiler (tested with \code{g++} 4.8.5 and higher,
  and \code{clang++} 3.3)
\item \cmake\footnote{See \url{http://cmake.org}.}, version \cmakeminversion
  or higher.
\item The \code{NLopt} library\footnote{See
  \url{https://nlopt.readthedocs.io/}.}, version \nloptminversion or
  higher.
\item Eigen library\footnote{See \url{http://eigen.tuxfamily.org}.},
  version \eigenminversion or higher
\item Boost libraries\footnote{See \url{http://www.boost.org}.},
  version \boostminversion or higher, specifically:
  \begin{itemize}
  \item[*] \code{Boost.Filesystem}
  \item[*] \code{Boost.Log}
  \end{itemize}
\item Our \ep library, which itself requires
  \begin{itemize}
  \item The \code{AlgLib} library
  \end{itemize}
\end{itemize}

On Ubuntu, they can be installed by
\begin{lstlisting}[language=bash]
$ sudo apt install libalglib-dev libnlopt-dev libeigen3-dev libboost-filesystem-dev libboost-log-dev
\end{lstlisting}

Furthermore, one of our example programs (\code{THDMIISNMSSMBC}) uses a potential built with \fs.
Our build script attempts to build \fs, but there are further dependencies; see the \fs manuals \cite{Athron:2014yba,Athron:2017fvs} or the \fs \href{https://github.com/FlexibleSUSY/FlexibleSUSY/blob/development/README.rst}{README file} for further details.

To visualize our results, our package includes plot scripts that require:
\begin{itemize}
\item \code{Python} and the external modules \code{scipy}, \code{numpy}, \code{matplotlib} and \code{pandas}.
\item \code{gnuplot}
\end{itemize}
On Ubuntu, they can be installed by
\begin{lstlisting}[language=bash]
$ sudo apt install gnuplot python-numpy python-scipy python-matplotlib python-pandas
\end{lstlisting}

Finally, for testing, we provide \bsmpt programs that require
\begin{itemize}
\item \code{libCMAES}. If this is absent, it is built automatically but this requires the \code{autoconf} and \code{libtool} build tools.
\end{itemize}

\subsection{Downloading and running \pt}\label{sec:download_etc}
\pt can be obtained by doing,
\begin{lstlisting}[language=bash]
git clone https://github.com/PhaseTracer/PhaseTracer
\end{lstlisting}
where the master branch will have the latest stable version, and the development branch will have the latest in-development features.

To build \pt on a UNIX-like system with the \make build system installed as the default build tool, run at the command line:

\begin{lstlisting}[language=bash]
$ cd PhaseTracer
$ mkdir build
$ cd build
$ cmake ..
$ make
\end{lstlisting}

The resulting library and executables are located in the \code{lib/} and \code{bin/} subdirectories of the main package directory, respectively. The built-in examples can be executed by running
\begin{lstlisting}[language=bash]
$ ./bin/run_1D_test_model -d
$ ./bin/run_2D_test_model -d
$ ./bin/scan_Z2_scalar_singlet_model -d
\end{lstlisting}
The results described in \secref{sec:examples} will be generated, including \figref{fig:2d}.

By default, the example \code{THDMIISNMSSMBC} is not compiled, as it needs \fs. To build the \code{THDMIISNMSSMBC} model, run at the command line:
\begin{lstlisting}[language=bash]
$ cd build
$ cmake -D BUILD_WITH_FS=ON ..
$ make
\end{lstlisting}
Then the example can be executed by
\begin{lstlisting}[language=bash]
$ ./bin/run_THDMIISNMSSMBCsimple
\end{lstlisting}
Finally, by default comparisons with \bsmpt are not compiled, but may be enabled by the \code{cmake -D BUILD_WITH_BSMPT=ON} flag.

\section{The physics problem}
\label{sec:PhysicalProblem}
Our aim is to trace the minima of the effective potential for some arbitrary BSM physics theory, and find the critical temperatures for possible phase transitions between them. The first task is to construct the effective potential.

\subsection{Constructing the effective potential}

The effective potential typically includes the following contributions,
\begin{equation}\label{Eq:FullPotential}
    V_\text{eff} = V^\text{tree} +  \Delta V^{\textrm{1-loop}} + \Delta V_T^{\textrm{1-loop}} + V_\text{daisy}.
\end{equation}
where $V^\text{tree}$ is the tree-level potential expressed in terms of \MSbar parameters.  The contribution $\Delta V^{\textrm{1-loop}}$ stands for the one-loop zero temperature corrections to the effective potential, which in the  $R_\xi$ gauge are given by~\cite{Patel:2011th},
\begin{equation}\label{eq:1LoopCorrection}
\begin{split}
\Delta V^{\textrm{1-loop}} = \frac{1}{64 \pi^2} \Bigg( & \sum_\phi n_\phi m_\phi^4(\xi) \left[\ln\left( \frac{m_\phi^2(\xi)}{Q^2}\right) - 3/2\right]\\
+ & \sum_V n_V m_V^4 \left[\ln\left(\frac{m_V^2}{Q^2}\right) - 5/6\right]\\
- & \sum_V \tfrac13 n_V (\xi m_V^2)^2 \left[\ln\left(\frac{\xi m_V^2}{Q^2}\right) - 3/2\right]\\
- & \sum_f n_f m_f^4 \left[\ln\left(\frac{m_f^2}{Q^2}\right) - 3/2\right]\Bigg).
\end{split}
\end{equation}
where the sums over $\phi$, $V$ and $f$ represent sums over the scalar, vector and fermion fields respectively and we have introduced the renormalization scale, $Q$, the field dependent \MSbar mass eigenstates of the considered BSM theory $m_i$, where there $i$ stands for the BSM fields which are summed over, and finally $n_i$, which gives the numbers of degrees of freedom for field $i$. With the potential constructed in this way we are assuming that the \MSbar parameters have been chosen or adjusted to fulfill the electroweak symmetry breaking conditions at the one-loop order, as would automatically be the case if one is obtaining them from a mass spectrum generator.

Please also note that if the parameters are input in a different scheme then one may need to modify these corrections, see e.g.\ section 3 of Ref.\ \cite{Martin:2001vx} for a review of the known one-loop corrections in the \MSbar, \DRbar \cite{Siegel:1979wq,Capper:1979ns} and $\overline{\textrm{DR}}^\prime$ \cite{Jack:1994rk} schemes in the Landau gauge.  One could also add counter-terms to the potential\footnote{By default, the counterterm potential is not included, but can be added to the potential, as described in \secref{sec:EffectivePotential_lib} where we discuss the code implementation.} in \eqref{Eq:FullPotential}, i.e.\ change $V_\text{eff} \rightarrow V_\text{eff} + V_\text{c.t.}$ modifying the \MSbar scheme.  This can be used to e.g.\ ensure that the one-loop zero temperature minimum matches the tree-level minimum and that the pole masses correspond to tree-level mass eigenstates.

The contribution $\Delta V_T^{\textrm{1-loop}}$ is the one-loop finite-temperature correction to the effective potential and in the $R_\xi$ gauge it is given by,
\begin{equation}\label{eq:thermal_one_loop}
\begin{split}
\Delta V_T^{\textrm{1-loop}} = \frac{T^4}{2 \pi^2} \Bigg[&
\sum_\phi n_\phi \JB\left(\frac{m_\phi^2(\xi)}{T^2}\right)
+ \sum_V n_V \JB\left(\frac{m_V^2}{T^2}\right)\\
- & \sum_V \tfrac13 n_V \JB\left(\frac{\xi m_V^2}{T^2}\right)
+ \sum_f n_f \JF\left(\frac{m_f^2}{T^2}\right)\Bigg] ,
\end{split}
\end{equation}
where $\JB$ and $\JF$ are the thermal functions,
\begin{equation}
\JBF(y^2) =
\pm \re
\int_0^{\infty}
		x^2 \ln
		\left(
			1 \mp \eu^{-\sqrt{x^2 + y^2}}
		\right)
dx.
\label{Eq:JBJF}
\end{equation}
The upper (lower) sign above applies for bosons (fermions).  Strong Boltzmann suppression occurs in these thermal functions for $m^2 \gg T^2$.

Finally {\it if} the Debye masses are provided then the daisy corrections can be included.  Following the Arnold-Espinosa approach we include these in an additive potential of the form,
\begin{equation}\label{eq:daisy}
V_{\text{daisy}} = -\frac{T}{12 \pi} \left(
\sum_\phi n_\phi \left[ \left(\overline m_\phi^2\right)^{\frac32} - \left(m^2_\phi\right)^{\frac32} \right] + \sum_V \tfrac13 n_V \left[\left(\overline m_V^2\right)^{\frac32} - \left(m^2_V\right)^{\frac32} \right]
\right),
\end{equation}
where in the first two terms we sum over the Higgs fields (including Goldstone bosons) and in the second two terms we sum over the massive gauge bosons. We use $\overline m^2$ to denote field-dependent mass eigenvalues that {\it include} Debye corrections.  If there are no Debye corrections to the masses this contribution will vanish. We always take the real part of the potential, which deals with the possibility of tachyonic masses in \eqref{eq:daisy}.  If Debye masses are not supplied by the user then this contribution will be neglected.

The effective potential presented is not gauge invariant~\cite{Garny:2012cg,Patel:2011th}. In our implementation, as described in \secref{sec:Implementing_Models}, we allow the user to choose any $R_\xi$ gauge for the calculation, allowing them to test variation in a variety of gauges.  In future releases we plan to add additional methods to allow an alternative gauge-independent calculation based on the procedure in Ref.\ \cite{Patel:2011th}. We also note that there are well-known problems associated with perturbative approaches near phase transitions~\cite{Linde:1980ts,Gross:1980br}. Nevertheless, lattice approaches are not currently a tractable alternative for phenomenological studies in arbitrary models of new physics.

\subsection{Critical temperature and transition strengths}

Having constructed a temperature dependent effective potential, we wish to trace its possible phases as the temperature changes.  Tracing phases means tracking the temperature dependence of the minima of the effective potential.  We will describe our method to do this in a separate section below.  The result of this tracing are the phases of the potential, that is the locations of all minima as functions of temperature. 

After we identified the phases of the potential we also wish to find critical temperatures at which the potential has degenerate minima.  We define the critical temperature, $T_C$, to be the temperature at which the potential has two degenerate minima,
\begin{equation}\label{eq:TcDefinition}
 V_\text{eff}(\minima{x}, T_C) = V_\text{eff}(\minima{x}^\prime, T_C) .
\end{equation}
Here $\minima{x}$ and $\minima{x}^\prime$ are two distinct minima in the field space, that is they are separated by a barrier.
Without loss of generality we can assume that the deepest vacuum of the system was $\minima{x}$ above $T_C$. In most cases, below $T_C$ the minima $\minima{x}^\prime$ becomes the deepest vacuum.
Just below the critical temperature, however, the system may remain at $\minima{x}$, i.e., in a false vacuum. Somewhat below $T_C$ the system may fluctuate over or tunnel through the barrier to the lower minimum, $\minima{x}^\prime$, called the true vacuum~\cite{Coleman:1977py, Callan:1977pt, Linde:1980tt}.

We also define the quantity which is useful for characterising the strength of the transition,
\begin{equation}
\gamma \equiv \frac{\Delta v (T_C)}{T_C} \qquad \textrm{where} \qquad \Delta v(T) = \sqrt{\sum_{i=1}^N (\minima{x_i} (T) -  \minima{x_i}^{\prime} (T))^2},
\end{equation}
and $N$ is the number of scalar fields considered.  By default all
scalar fields in the potential are considered in the quadratic sum
defining $\Delta v(T)$, but as described later this may also be restricted,
allowing, for example, one to output $\gammaEW \equiv (v_\text{EW}
(T_c) - v_\text{EW}^\prime (T_C)) / T_C $ which may be more relevant
for electroweak baryogenesis.

The strength of a first order phase transition can be a useful
heuristic for assessing whether the phase transition is strong enough
for a successful electroweak baryogenesis mechanism, with
$\gammaEW > 1$ being a commonly applied rule of thumb.
Similarly the strength of a first order phase transition also has
an impact on the detectability of gravitational waves that are
generated from the transition.

However it is important to note that for a full calculation of the
baryon asymmetry of the Universe or gravitational wave spectra many
other steps are required, which \pt alone does not take care of.  For
example another very important quantity, the so-called {\it nucleation
  temperature} can only be calculated by using \pt in combination with
a cosmological bounce solver, such as \bp \cite{Athron:2019nbd}, which
calculates the bounce action for the potential at a given
temperature. A first order transition proceeds via bubbles of broken
phase nucleating in space-time.  If these bubbles prove to be stable
and grow, somewhat below $T_C$ there will be a nucleation temperature,
$T_n$, at which the mean number of bubbles nucleated within the
relevant space-time volume is one.  For the early Universe the time of
nucleation can be written as
\begin{equation}
  \int_{t_c}^{t_n} dt \frac{\Gamma(t)}{H(t)^3} = \int_{T_n}^{T_C} \frac{dT}{T} \frac{\Gamma(T)}{H(t)^4} = 1 ,
\end{equation}
where $H$ is the Hubble parameter and $\Gamma(T)$ is the nucleation
rate per unit volume, which may be expressed as, $\Gamma(T)=
A(T)e^{-{S_E}/{T}}$ where $S_E$ is the bounce action that can be
obtained from a bounce solver.  Ultimately, the physical problem
becomes computing observable quantities --- such as gravitational wave
signatures or the baryon asymmetry of the Universe --- using some of
the above thermal properties of the transition.  For reviews on how to
do such calculations see
e.g.\ Refs.\ \cite{Trodden:1998ym,Quiros:2007zz,Morrissey:2012db,White:2016nbo,Mazumdar:2018dfl,Caprini:2019egz}.
While \pt does not provide all of these quantities, the outputs of
\pt, namely the phase structure and its temperature evolution, the
critical temperatures and the strengths of the transitions, are
important ingredients in such calculations and have quite broad
utility.

\subsection{Locating minima of the potential}

We locate all minima of the potential at $T = 0$ and $T = 1000 \gev$ (which may be changed by \code{PhaseTracer::set_t_low} and \code{PhaseTracer::set_t_high}). To do so, we first generate a set of guesses by uniform sampling. We polish the guesses using local minimization and keep the unique minima. There are obviously more sophisticated techniques for global optimization that we do not implement. Nonetheless our method is fast and we have tested that it gives reliable results (see \secref{sec:examples}).  If the performance of our implementation is a bottleneck or fails, we advise that a user implements techniques tailored to their particular physics problem or supplies guesses for the locations of the minima (\code{PhaseTracer::set_guess_points}). After locating the minima at low and high temperature, we trace the high-temperature minima down in temperature, and the low-temperature minima up.

\subsection{Tracing a minima}

After identifying the minima, we trace them with temperature using \code{PhaseFinder::trace_minimum} with an estimate of the derivative of the fields with respect to temperature,
\begin{equation}\label{eq:dx}
\Delta \minima{x} = \diff{\minima{x}}{T} \Delta T.
\end{equation}
Note that the minima, $\minima{x}$, carries an implicit dependence on temperature, i.e., by $\minima{x}$ we denote the trajectory of a minimum with temperature. We find the derivative by noting that
\begin{equation}
\diff{}{T} \diffp{V}{{x_i}} = \diffp{V}{{x_i}{x_j}} \diff{x_j}{T} + \diffp{V}{{x_i}{T}}
\end{equation}
must vanish when evaluated along the trajectory of a minima, $\minima{x}$, since by definition of the trajectory
\begin{equation}
\left.\diffp{V}{{x_i}}\right|_{\minima{x}} = 0
\end{equation}
for all temperatures, $T$, and all fields, $x_i$. Thus we find the derivative in \refeq{eq:dx} by solving
\begin{equation}\label{eq:dx_min_dt}
\left.\diffp{V}{{x_i}{x_j}}\right|_{\minima{x}} \diff{\minima{x_j}}{T} = - \left.\diffp{V}{{x_i}{T}}\right|_{\minima{x}}.
\end{equation}
This is implemented in \code{PhaseFinder::dx_min_dt}. The derivatives of the potential with respect to the fields and temperature are found by a finite difference method. We use it to predict the minima at temperature $T_1 = T_0 + \Delta T$,
\begin{equation}
x_{\text{predict}} \equiv \minima{x_0} + \Delta \minima{x_0}.
\end{equation}
We polish this prediction with the \code{PhaseFinder::find_min} method to find $\minima{x_1}$. Let us furthermore define a postdiction for $\minima{x_0}$ in a similar manner,
\begin{equation}
x_{\text{postdict}} \equiv \minima{x_1} - \Delta \minima{x_1}.
\end{equation}

Thus we trace a minimum in steps of $\Delta T$ by guessing the minimum at temperature $T_1 = T_0 + \Delta T$, polishing that guess, and iterating. We interrupt tracing if the step in temperature $\Delta T$ is smaller than a user-specified level (changed by \code{PhaseTracer::set_dt_min}) and if any of the following occurs:
\begin{itemize}
\item We encounter a discontinuous jump in the minimum.

We define a discontinuous jump by a significant difference between the minima, $\minima{x_1}$ and $\minima{x_0}$, and a significant difference between a minima and its expected location, that is, between $\minima{x_1}$ and $x_{\text{predict}}$ or between $\minima{x_0}$ and $x_{\text{postdict}}$. A significant difference is defined by the absolute and relative tolerances, \code{x_abs_jump} and \code{x_rel_jump}, that we use in \refeq{eq:compare_float}. Their default values are shown in \tabref{tab:setting-phases}.

\item The minimum goes out of bounds or into a forbidden region of field space.

\item The Hessian is not positive semi-definite at what should be a minima. There is a possibility that the \code{PhaseFinder::find_min} method may return a saddle point, instead of a local minimum. Thus if the Hessian at $x_1$ has any negative eigenvalues then the phase ends.

\item The determinant of the Hessian matrix is zero, as this indicates that there is probably a transition. Note that this criteria (disabled by \code{PhaseTracer::set_check_hessian_singular(false)}) can separate phases by first- and second-order transitions (without it, they are separated only by first-order ones).

\end{itemize}
If tracing isn't interrupted, we stop once we reach the desired temperature. If a jump or the Hessian indicated that a phase ended, we check where the minimum went by finding and tracing the new minimum. For performance, before tracing a minimum, we check that it doesn't belong to an existing known phase. This enables our method to efficiently trace intermediate phases that do not exist at the low or high temperature.

At each iteration, we re-evaluate the step size. If we find reason for interrupting tracing, we reduce the step size by a factor of two. If the guess of the minima, $x_{\text{guess}}$ lay far away from the real minima, we reduce the step size by a factor of two; if not, we increase the step size by a factor of two.
We define far away by the relevant absolute and relative tolerances, \code{x_abs_identical} and \code{x_rel_identical}, that we use in \refeq{eq:compare_float}. Their default values are shown in \tabref{tab:setting-phases}.

Our strategy for tracing a minima relies on local optimization of a reasonable guess, found from \refeq{eq:dx}. By default, we use a Nelder-Mead variant called subplex~\cite{Rowan90functionalstability} implemented in \code{Nlopt}~\cite{nlopt} as \code{LN_SBPLX}. This can fail. If we are tracing a second-order phase transition, the minimum won't change smoothly and thus our guess, based upon a first order Taylor expansion, may be poor. If the potential is pathological in the vicinity of the local minimum --- e.g., particularly flat --- the local optimization may fail even with a reasonable guess. Unfortunately, second-order transitions can present both problems simultaneously. In these cases, a phase may be incorrectly broken into two pieces that are discovered separately by the algorithm, since due to the discontinuity, tracing up and down can yield different results. The field values at the joint may be unreliable and the phases could slightly overlap. 

To remedy this, note that we can predict first- and second-order transitions by singularities in the Hessian matrix in \refeq{eq:dx_min_dt}. For second-order transitions, a singular Hessian matrix means that there may be multiple solutions for the change in minimum with respect to temperature. The presence of multiple solutions for $\idiff{\minima{x}}{T}$ indicates a possible jump discontinuity in $\idiff{\minima{x}}{T}$ at that temperature and thus a second-order transition. In cases in which a phase ends by a first order transition, on the other hand, two extrema --- a minima and a barrier --- merge into a double root. The Hessian matrix must vanish at a double root. Let us denote the magnitude of the smallest eigenvalue of the Hessian matrix at $\minima{x}$ at temperature $T$ by $\minima{\lambda_T}$ and at zero temperature by $\minima{\lambda_0}$. We judge the Hessian matrix to be singular if
\begin{equation}\label{eq:hessian_singular}
\minima{\lambda_T} < \epsilon_{\textrm{hess}} \cdot \minima{\lambda_0}.
\end{equation}
where $\epsilon_{\textrm{hess}}$ is a relative tolerance for this, stored in a member of the \code{PhaseFinder} class of \pt, \code{double hessian_singular_rel_tol}.  This has a default value of $0.01$ which we recommend for most cases, but this may be reset by the user, via the method \code{set_hessian_}\code{singular_rel_tol(double tol)}.  We use the smallest eigenvalue of the Hessian at zero temperature as an appropriate numerical scale, as if \refeq{eq:hessian_singular} is satisfied it means that there must be cancellations in the Hessian matrix caused by the finite-temperature corrections at that particular temperature. This partially alleviates the potential problems discussed in the previous paragraph.

\subsection{Dealing with discrete symmetries}\label{sec:symmetries}

Discrete symmetries are common in scalar potentials. In a model with $n$ discrete symmetries, each phase is potentially duplicated $2^n$ times (though it may transform trivially under some transformations). The duplicates, however, cannot be ignored. Let us denote two phases by $P$ and $Q$, which transform to $P^\prime$ and $Q^\prime$ by a discrete symmetry of the model. Suppose that at some critical temperature a transition from $P \to Q$ is possible. The discrete symmetry means that the transitions $P \to Q$ and $P^\prime \to Q^\prime$ are equivalent, and that the transitions $P \to Q^\prime$ and $P^\prime \to Q$ are equivalent, but it does not mean that e.g., $P \to Q$ and $P \to Q^\prime$ are equivalent. In fact, in the presence of $n$ discrete symmetries, each transition belongs to a set of up to $2^n$ inequivalent ``cousin'' transitions.

Thus, to account for discrete symmetries efficiently, we allow a user to specify them if they want in the virtual \code{std::vector<Eigen::VectorXd> apply_symmetry(Eigen::VectorXd phi)} function. This insures that only one copy of a phase is traced, but all copies related by discrete symmetries are taken into account when calculating possible transitions. The function should return the result of each discrete symmetry on the set of fields, i.e., for $n$ symmetries the returned \code{std::vector} should have size $n$. We do not address global or local continuous symmetries, leaving it to a user to gauge fix them appropriately in their potential if they wish. See \secref{sec:2D_test_model} for an example.

\subsection{Finding possible first order transitions}\label{sec:finding_FOPTs}

Having identified the phases, we check for possible first order transitions (FOPTs) between them. For every pair of phases, we find the temperature interval in which they coexist. If the interval is non-vanishing and if the difference in potential between the two phases, $\Delta V$, changes sign over the interval, we look for a critical temperature with a root-finding algorithm. By default, we assume there can be no more than one critical temperature between any two phases. This can be switched off by \code{TransitionFinder::set_assume_only_one_transition(false)} and specifying the minimum separation between critical temperatures by e.g., \code{TransitionFinder::set_separation(1.5)}. With these settings, we would look for critical temperatures in every $1.5\gev$ interval in temperature in which the phases coexisted. The current default value for this setting is $1$ GeV, based on experience with phase transitions that we have tested \pt on, i.e.\ the examples discussed in \secref{sec:examples}.

To check the validity of critical temperatures by eye, in \code{potential_line_plotter.hpp} we provide a plotting routine that shows the potential along a straight line between the true and false minima at the critical temperature. We suggest that it is called as e.g., \code{PhaseTracer::potential_line_plotter(EffectivePotential::Potential &P, tf.get_transitions())}. For a genuine critical temperature, we should see degenerate minima separated by a barrier.   We warn readers that we encountered cases in which numerical artefacts in the potential --- e.g., small jump discontinuities in potentials that were constructed in a piece-wise manner --- were located by our code and mistaken for minima. Thus, we advise checking a few transitions by eye if the potential contains any possible numerical pathologies.

\section{\pt structure}\label{sec:structure}

\begin{figure}[h!]
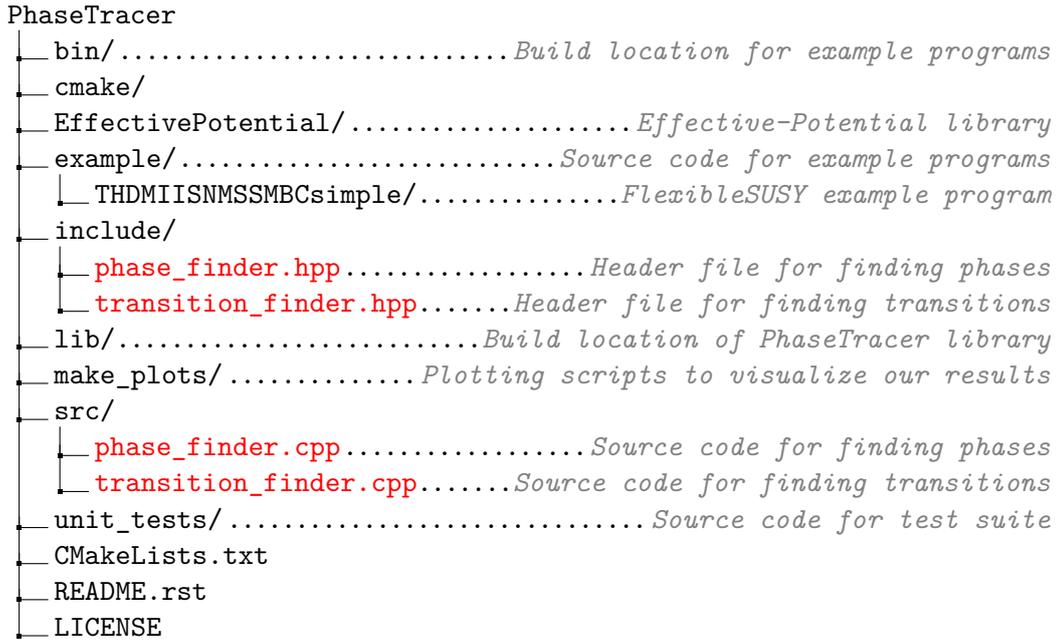

\centering
\framebox[\textwidth]{%
\begin{minipage}{0.95\textwidth}
\vspace{2mm}
	\dirtree{%
		.1 PhaseTracer.
		.2 bin/\DTcomment{Build location for example programs}.
		.2 cmake/.
    .2 EffectivePotential/\DTcomment{Effective-Potential library}.
    .2 example/\DTcomment{Source code for example programs}.
  		.3 THDMIISNMSSMBCsimple/\DTcomment{\fs example program}.
		.2 include/.
  		.3 \DThighlight{phase\_finder.hpp}\DTcomment{Header file for finding phases}.
      .3 \DThighlight{transition\_finder.hpp}\DTcomment{Header file for finding transitions}.
    .2 lib/\DTcomment{Build location of \pt library}.
    .2 make\_plots/\DTcomment{Plotting scripts to visualize our results}.
		.2 src/.
  		.3 \DThighlight{phase\_finder.cpp}\DTcomment{Source code for finding phases}.
		  .3 \DThighlight{transition\_finder.cpp}\DTcomment{Source code for finding transitions}.
		.2 unit\_tests/\DTcomment{Source code for test suite}.
		.2 CMakeLists.txt.
		.2 README.rst.
		.2 LICENSE.
	}
\vspace{2mm}
\end{minipage}
}
\caption{Outline of the structure of \pt. We highlight the most important files in red and do not show all files.}\label{fig:DirectoryTree}
\end{figure}

The problem is divided into three steps: constructing the potential, finding phases and checking for critical temperatures between them. They are performed by the \ep library, and the \code{PhaseFinder} and \code{TransitionFinder} classes, respectively. The latter are implemented in the \code{phase_finder.\{hpp,cpp\}} and \code{transition_finder.\{hpp,cpp\}} files in the source code, respectively. The structure is illustrated in \figref{fig:DirectoryTree}.

\section{Implementing new models and running the code}
\label{sec:Implementing_Models}
\subsection{The \ep library}
\label{sec:EffectivePotential_lib}
New models are implemented in our \ep library, which provides pure virtual classes that represent a tree-level (\code{EffectivePotential::Potential}) and a one-loop effective potential (\code{EffectivePotential::OneLoopPotential}). The latter automatically includes one-loop zero- and finite temperature corrections, as well as daisy corrections.

\pt comes with example models that publicly inherit from the pure virtual \code{EffectivePotential::Potential} class. The pure virtual classes make it easy to implement a new model. For example, we may implement the simple 1D model in \secref{sec:1D_test_model} as,
\newline\begin{minipage}{\linewidth}
\begin{lstlisting}
#include "potential.hpp"
#include "pow.hpp"

namespace EffectivePotential {
// Publicy inherit from the EffectivePotential class
class OneDimModel : public Potential {
  public:
    // Implement our scalar potential - this is compulsory
    double V(Eigen::VectorXd x, double T) const override {
      return (0.1 * square(T) - square(100.)) * square(x[0])
             - 10. * cube(x[0]) + 0.1 * pow_4(x[0]);
    }
    // Declare the number of scalars in this model - this is compulsory
    size_t get_n_scalars() const override { return 1; }
    // Look at x >= 0 - this is optional
    bool forbidden(Eigen::VectorXd x) const override { return x[0] < -0.1;}
};
}
\end{lstlisting}
\end{minipage}
We have overriden two pure virtual methods: \code{V}, the scalar potential as a function of the fields and the temperature, and \code{get_n_scalars}, the number of scalar fields in this problem. We have furthermore overridden the virtual method \code{forbidden}, which ensures that our field always has non-negative values. In the implementation of this model packaged with \pt in \code{models/1D_analytic_test_model.hpp} we furthermore implement analytic derivatives of the potential. In this simple example we are not including any additional perturbative corrections.

To include such perturbative corrections one can instead implement one-loop effective potentials through the pure virtual \code{EffectivePotential::OneLoopPotential} class. For example, we may implement the 2D example in \secref{sec:2D_test_model}  as
\begin{lstlisting}
#include <vector>
#include "one_loop_potential.hpp"
#include "pow.hpp"

namespace EffectivePotential {
  class TwoDimModel : public OneLoopPotential {
   public:
    double V0(Eigen::VectorXd phi) const override {
      return 0.25 * l1 * square(square(phi[0]) - square(v))
             + 0.25 * l2 * square(square(phi[1]) - square(v))
             - square(mu) * phi[0] * phi[1];
      }

    std::vector<double> get_scalar_masses_sq(Eigen::VectorXd phi,
                                             double xi) const override {
      const double a = l1 * (3. * square(phi[0]) - square(v));
      const double b = l2 * (3. * square(phi[1]) - square(v));
      const double A = 0.5 * (a + b);
      const double B = std::sqrt(0.25 * square(a - b) + pow_4(mu));
      const double mb_sq = y1 * (square(phi[0]) + square(phi[1])) + y2 * phi[0] * phi[1];
      return {A + B, A - B, mb_sq};
   }

  std::vector<double> get_scalar_dofs() const override { return {1., 1., 30.}; }
  size_t get_n_scalars() const override { return 2; }

  std::vector<Eigen::VectorXd> apply_symmetry(Eigen::VectorXd phi) const override {
    return {-phi};
  }

 private:
  const double v = 246.;
  const double m1 = 120.;
  const double m2 = 50.;
  const double mu = 25.;
  const double l1 = 0.5 * square(m1 / v);
  const double l2 = 0.5 * square(m2 / v);
  const double y1 = 0.1;
  const double y2 = 0.15;
};
}
\end{lstlisting}

This time we have overridden five methods of the pure virtual \code{OneLoopPotential} class to define our model: \code{V0}, the tree-level potential; \code{get_scalar_masses_sq}, the field-dependent scalar squared masses; \code{get_scalar_dof}, the numbers of degrees of freedom for the scalars; \code{get_n_scalars}, the number of scalar fields (2) in this problem; and lastly, \code{apply_symmetry}, which  returns the result of the model's discrete symmetry, $\phi \to -\phi$, on the fields.

The \code{OneLoopPotential} class itself implements methods for the one-loop zero-temperature and finite-temperature corrections to the potential. Fermion and vector contributions are calculated if the methods \code{get_\{fermion/vector\}_masses_sq} and  \code{get_\{fermion/vector\}_dofs} are implemented. By default, it works in the $\xi=0$ (Landau) gauge, but this may be changed to any $R_\xi$ gauge by the \code{OneLoopPotential::set_xi} method. Note, though, that it is up to the user to correctly implement the $\xi$-dependence of the scalar masses if they depart from $\xi = 0$. Daisy contributions are added using the Arnold-Espinosa~\cite{Arnold:1992rz} method if Debye masses are supplied by implementing \code{get_\{fermion/vector\}_debye_sq}. Furthermore, counter-terms can be added to the potential by overriding the virtual method \code{counter\_term}.

The only methods that must be overriden (i.e., that are pure virtual), however, are the tree-level potential and the number of scalar fields. The \code{get_scalar_dof} and \code{get_scalar_mass_sq} methods aren't compulsory --- they default to one degree of freedom per scalar field and numerical estimates of the eigenvalues of the Hessian matrix of the tree-level potential, respectively. If you change the numbers of degrees of freedom, it is vital to change \code{get_scalar_mass_sq} too, as the order of the eigenvalues when found numerically won't be stable and won't correspond correctly to the intended degrees of freedom per scalar field. For this reason and for accuracy, wherever possible it is wise to implement derivatives and eigenvalues analytically. By default, there are no fermions or vectors in the model and the \code{apply_symmetry} method trivially returns the coordinates with no changes.

The  \code{OneLoopPotential} class automatically includes one-loop thermal corrections to the potential in~\eqref{eq:thermal_one_loop}. This requires an implementation of the thermal functions in \eqref{Eq:JBJF}. We interpolated them from a set of look-up tables generated from \cosmo.

\subsection{Running \pt}

\pt may be called in an example program as follows (using the \code{EffectivePotential::OneDimModel} as an example),
\newline\begin{minipage}{\linewidth}
\begin{lstlisting}
#include <iostream>
#include "models/1D_test_model.hpp"
#include "phase_finder.hpp"
#include "transition_finder.hpp"

int main() {
  EffectivePotential::OneDimModel model;  // Construct the 1D model

  PhaseTracer::PhaseFinder pf(model);  // Construct the PhaseFinder
  pf.find_phases();  // Find the phases
  std::cout << pf;  // Print information about the phases

  PhaseTracer::TransitionFinder tf(pf);  // Construct the TransitionFinder
  tf.find_transitions();  // Find the transitions
  std::cout << tf;  // Print information about the transitions
  return 0;
}
\end{lstlisting}
\end{minipage}
For a new model, \code{EffectivePotential::OneDimModel} should be replaced with the name of the new model. The line \code{std::cout << pf;} produces output about the phases with the format:

\begin{lstlisting}[language=bash]
found 2 phases

=== phase key = 0 ===
Maximum temperature = 1000
Minimum temperature = 33.1513
Field at tmax = [1.48461e-05]
Field at tmin = [1.52403e-05]
Potential at tmax = 2.20185e-05
Potential at tmin = 2.29965e-09
Ended at tmax = Reached tstop
Ended at tmin = Jump in fields indicated end of phase

=== phase key = 1 ===
Maximum temperature = 61.7437
Minimum temperature = 0
Field at tmax = [37.832]
Field at tmin = [81.1597]
Potential at tmax = 65886.6
Potential at tmin = -1.66587e+06
Ended at tmax = Jump in fields indicated end of phase
Ended at tmin = Reached tstop
\end{lstlisting}
We see the number of phases found, and for each phase, the minimum and maximum temperature, the corresponding field values and potential, and an explanation about why the phase ended. In this case, the first phase ended at $T = 1000\gev$ because that was the highest temperature under consideration (see \code{PhaseFinder::set_t_high}) and at $T \approx 33\gev$ because the fields made a discontinuous jump. Each phase is numbered by a key, e.g.\ \code{=== phase key = 0 ===}.

The line \code{std::cout << tf;} produces output about the transitions with the format:
\begin{lstlisting}[language=bash]
found 1 transition

=== transition from phase 0 to phase 1 ===
true vacuum = [50.0003]
false vacuum = [1.09314e-05]
changed = [true]
TC = 59.1608
gamma = 0.845159
delta potential = 0.00117817
\end{lstlisting}
We see the number of potential first order phase transitions, and for each one, the keys of the false and true phases (i.e., \code{=== transition from phase 0 to phase 1 ===}), followed by the true and false vacua at the critical temperature, information about which elements of the field changed in the transition, the critical temperature, transition strength $\gamma$ ( or $\gammaEW$ if \code{TransitionFinder::set_n_ew_scalars} is used) and difference in potential between the true and false vacua at the critical temperature. The latter serves as a sanity check: it should of course be close to zero.

\subsubsection{The \code{Phase} and \code{Transition} objects}

The objects containing this information may be accessed though \code{PhaseFinder::get_phases()} and \code{TransitionFinder::get_transitions()}, which return a \code{std::vector} of \code{Phase} and \code{Transition} objects, respectively. The \code{Phase} object is a \code{struct} containing, amongst other things, the attributes:
\begin{itemize}
\item \code{std::vector<Eigen::VectorXd> X} --- The field values, $\minima{x}$, through the phase
\item \code{std::vector<double> T} --- The temperature, $T$, through the phase
\item \code{std::vector<Eigen::VectorXd> dXdT} --- The change in minima with respect to temperature, $\idiff{\minima{x}}{T}$, through the phase
\item \code{std::vector<double> V} --- The potential, $V(\minima{x}, T)$, through the phase
\item \code{phase_end_descriptor end_low} --- The reason why the phase ended at the lowest temperature
\item \code{phase_end_descriptor end_high} --- The reason why the phase ended at the highest temperature
\end{itemize}
The \code{std::vector} attributes are all ordered in ascending temperature. The \code{Transition} object, on the other hand, contains
\begin{itemize}
\item \code{double TC} --- The critical temperature, $T_C$
\item \code{Phase true_phase} --- The phase associated with the true vacuum
\item \code{Phase false_phase} --- The phase associated with the false vacuum
\item \code{Eigen::VectorXd true_vacuum} --- The true vacuum at the critical temperature
\item \code{Eigen::VectorXd false_vacuum} --- The false vacuum at the critical temperature
\item \code{size_t key} --- The key indicating unique cousin transition.
\end{itemize}

 For example, we may retrieve and print information about the first transition by
\begin{lstlisting}
auto transitions = tf.get_transitions();
std::cout << transitions[0].TC;  // Print just the critical temperature
std::cout << transitions[0].true_phase;  // Summary of true phase
std::cout << transitions[0].true_phase.key;  // Key of the true phase
std::cout << transitions[0];  // Summary information about the transition
\end{lstlisting}
which would tell us the critical temperature of the first transition, and then print a summary of information about the true phase, the key corresponding to the true phase, and finally a summary of information about the transition.

\subsubsection{Settings for finding phases and transitions}

There are many adjustable settings that control the behavior of \code{PhaseFinder} and \allowbreak \code{TransitionFinder} objects, listed in \tabref{tab:setting-phases} and \tabref{tab:setting-transitions}, respectively. They are altered by setters, i.e.,  \code{set_\{setting_name\}} methods, and read by \code{get_\{setting_name\}} methods. The detailed usage of settings is introduced in \secref{sec:PhysicalProblem} when we describe our algorithms of tracing phases and finding critical temperatures.

The attribute \code{n_ew_scalars} in both \code{PhaseFinder} and \code{TransitionFinder} is the number of scalar fields charged under electroweak symmetry, which we assume are the first \code{n_ew_scalars} scalar fields. By default \code{PhaseFinder} checks that the zero-temperature vacuum agrees with $v = 246\gev$ only if a user sets \code{n_ew_scalars} to a non-zero value. By default \code{TransitionFinder} calculates the transition strength $\gamma$ using all scalar fields; it calculates $\gammaEW$ only if a user sets \code{n_ew_scalars}.

When comparing floating point numbers, we typically check relative and absolute differences, i.e., our checks are of the form,
\begin{equation}\label{eq:compare_float}
|a - b| \le \text{\code{abs\_tol}} + \text{\code{rel\_tol}} \cdot \max(|a|, |b|).
\end{equation}
When tracing phases in temperature, we restrict the largest and smallest possible step sizes. The largest permissible step-size \code{dt_max} is the minimum of \code{dt_max_abs} and the temperature interval multiplied by \code{dt_max_rel}. Similarly, the smallest permissible step-size \code{dt_min} is the maximum one found from the relative and absolute settings.

\begin{table}[]
\small{
\begin{tabularx}{\textwidth}{llX}
\toprule
\codetab{setting\_name}        & Default value      & Description                                   \\
\midrule
\codetab{seed}                     & -1        & Random seed (if negative, use system clock)                \\
\codetab{upper_bounds}             & $1600\gev$         & Upper bounds on fields, can be altered to a \codetab{std::vector<double>} of \codetab{n_scalars}-dimension \\
\codetab{lower_bounds}             & $-1600\gev$        & Lower bounds on fields, same as above     \\
\codetab{guess_points}             & \codetab{\{\}}     & Guesses for locations of minima           \\
\codetab{n_test_points}            &  100               & Number of generated guesses with which to find all minima before tracing them \\
\codetab{t_low}                    & $0\gev$            & Lowest temperature to consider ($T_l$)    \\
\codetab{t_high}                   & $1000\gev$         & Highest temperature to consider ($T_h$)   \\
\codetab{check_vacuum_at_high}     & \codetab{true}     & Check whether there is a unique vacuum at $T_h$ \\
\codetab{check_vacuum_at_low}      & \codetab{true}     & Check whether deepest vacuum at $T_l$ agrees with that expected  \\
\codetab{n_ew_scalars}             & 0                  & Number of scalar fields charged under electroweak symmetry, used for checking vacuum at zero-temperature \\
\codetab{v}                        & $246\gev$          & Zero-temperature vacuum expectation value of electroweak charged scalars \\
\codetab{x_abs_identical}          & $1\gev$            & Absolute error below which field values are considered identical \\
\codetab{x_rel_identical}          & $1\mathrm{e}{-3}$  & Relative error below which field values are considered identical \\
\codetab{x_abs_jump}               & $0.5\gev$          & Absolute change in field that is considered a jump \\
\codetab{x_rel_jump}               & $1\mathrm{e}{-2}$  & Relative change in field that is considered a jump \\
\codetab{find_min_algorithm}       & \codetab{nlopt::LN\_SBPLX} & Algorithm for finding minima      \\
\codetab{find_min_x_tol_rel}       & $1\mathrm{e}{-4}$  & Relative precision for finding minima     \\
\codetab{find_min_x_tol_abs}       & $1\mathrm{e}{-4}\gev$  & Absolute precision for finding minima \\
\codetab{find_min_max_f_eval}      & $1\mathrm{e}{+6}$  & Maximum number of function evaluations when finding minimum \\
\codetab{find_min_min_step}        & $1\mathrm{e}{-4}$  & Minimum step size for finding minima      \\
\codetab{find_min_max_time}        & 5 minutes          & Timeout for finding a minima              \\
\codetab{find_min_trace_abs_step}  & $1\gev$            & Initial absolute step size when tracing a minimum   \\
\codetab{find_min_locate_abs_step} & $1\gev$            & Initial absolute step size when finding a minimum   \\
\codetab{dt_start_rel}             & 0.01               & The starting step-size relative to $T_h-T_l$        \\
\codetab{t_jump_rel}               & 0.005              & The jump in temperature from the end of one phase to the temperature at which we try to trace a new phase. \\
\codetab{dt_max_abs}               & $50.\gev$          & The largest absolute step-size in temperature       \\
\codetab{dt_max_rel}               & 0.25               & The largest relative step-size in temperature       \\
\codetab{dt_min_rel}               & $1\mathrm{e}{-7}$  & The smallest relative step-size in temperature      \\
\codetab{dt_min_abs}               & $1\mathrm{e}{-10}\gev$  & The smallest absolute step-size in temperature \\
\codetab{hessian_singular_rel_tol} & $1\mathrm{e}{-2}$  & Tolerance for checking whether Hessian was singular \\
\codetab{linear_algebra_rel_tol}   & $1\mathrm{e}{-3}$  & Tolerance for checking solutions to linear algebra  \\
\codetab{trace_max_iter}           & $1\mathrm{e}{+5}$  & Maximum number of iterations when tracing a minimum \\
\codetab{phase_min_length}         & $0.5\gev$          & Minimum length of a phase in temperature            \\
\bottomrule
\end{tabularx}}
\caption{Adjustable settings that control the behavior of a \code{PhaseFinder} object. Each setting has \code{get_\{setting_name\}} and \code{set_\{setting_name\}} getter and setter methods.}
  \label{tab:setting-phases}
\end{table}

\begin{table}[]
\begin{tabularx}{\textwidth}{llX}
\toprule
\codetab{setting_name}               & Default value          & Description                                                                     \\
\midrule
\codetab{n_ew_scalars}               & -1                     & Number of scalar fields charged under electroweak symmetry
                                                                that contribute to $\gamma$. If \codetab{-1}, include all fields in $\gamma$   \\
\codetab{TC_tol_rel}                 & $1\mathrm{e}{-4}$      & Relative precision in critical temperature                                               \\
\codetab{max_iter}                   & 100                    & Maximum number of iterations when finding a critical temperature                         \\
\codetab{trivial_rel_tol}            & $1\mathrm{e}{-3}$      & Relative tolerance for judging whether a transition was trivial as fields did not change \\
\codetab{trivial_abs_tol}            & $1\mathrm{e}{-3}\gev$  & Absolute tolerance for judging whether a transition was trivial as fields did not change \\
\codetab{assume_only_one_transition} & \codetab{true}         & Assume at most one critical temperature between two phases                               \\
\codetab{separation}                 & $1\gev$                & Minimum separation between critical temperatures                                         \\
\bottomrule
\end{tabularx}
\caption{Adjustable settings that control the behavior of \code{TransitionFinder} object.}
  \label{tab:setting-transitions}
\end{table}

\subsection{Plotting results}

The header \code{include/phase_plotter.hpp} provides a function \code{phase_plotter(PhaseTracer::TransitionFinder tf, std::string prefix = "model")} that takes a \code{TransitionFinder} object and plots the phases, using \code{prefix} to construct file names for the resulting plots and data files:
\begin{itemize}
\item  \code{\{prefix\}.dat} --- Data file containing phases and transitions in a particular format
\item \code{phi_1_phi_2_\{prefix\}.pdf} etc --- The phases and transitions on plane of the first and second field, e.g., the left panel of \figref{fig:2d}
\item \code{phi_T_\{prefix\}.pdf} --- The minima, $\minima{x}$, versus temperature, $T$, for every phase, e.g., the right panel of \figref{fig:2d}
\item \code{V_T_\{prefix\}.pdf}  --- The potential, $V(\minima{x}, T)$, versus temperature, $T$, for every phase.
\end{itemize}
The plots are produced by the script \code{make_plots/phase_plotter.py}. To make plots with \LaTeX\ fonts, \code{export MATPLOTLIB_LATEX}. Note that in models with discrete symmetries supplied by a user, the plots do not show cousin transitions (see \secref{sec:symmetries}).

We furthermore provide \code{include/potential_line_plotter.hpp} to check critical temperatures by plotting the minima between the true and false vacuum at the critical temperature --- see \secref{sec:finding_FOPTs} for further details.

\section{Examples/Comparisons with existing codes and analytic solutions}
\label{sec:examples}

\subsection{One-dimensional test model}\label{sec:1D_test_model}

\begin{figure}[t]
\centering
\includegraphics[height=.34\textwidth]{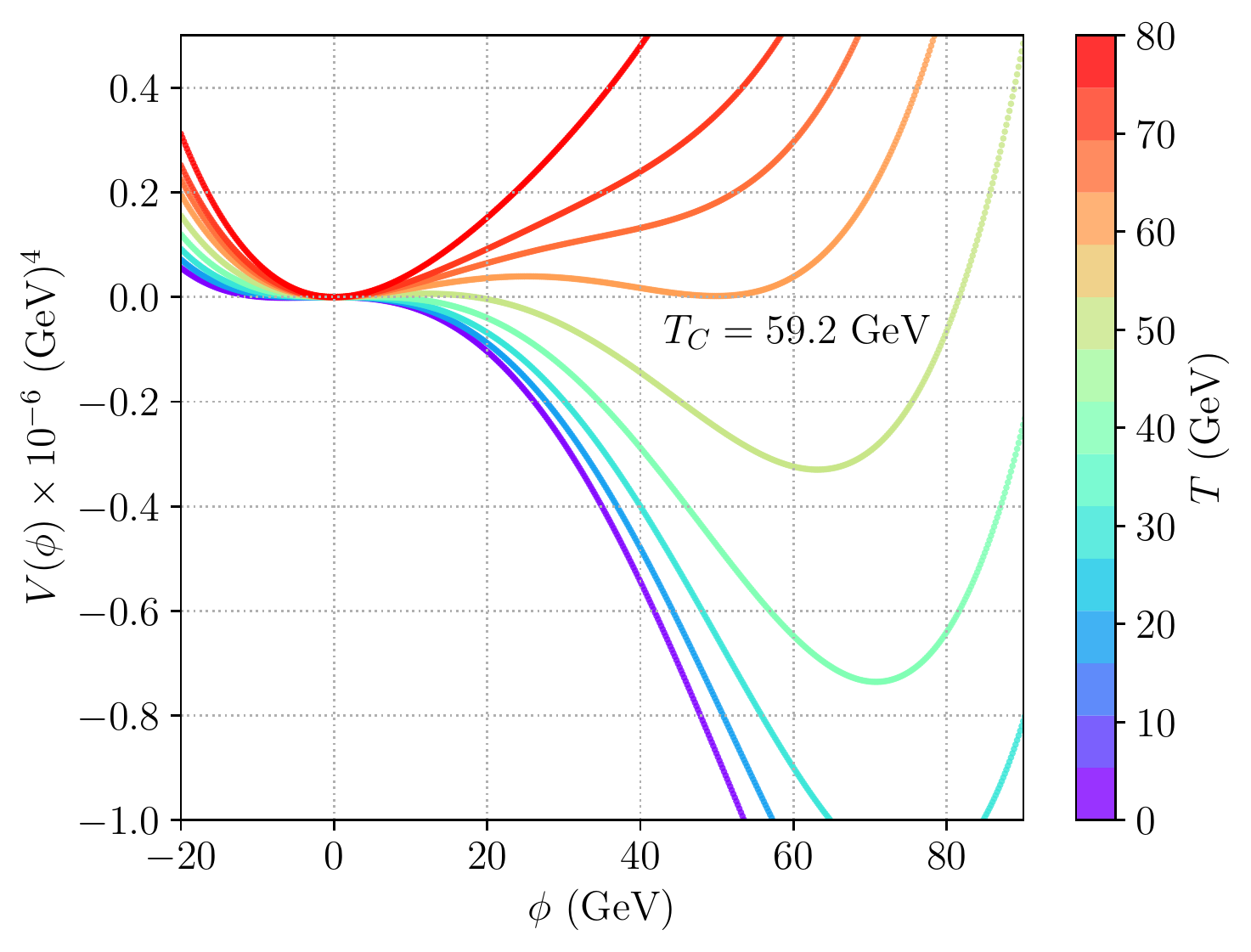}
\includegraphics[height=.34\textwidth]{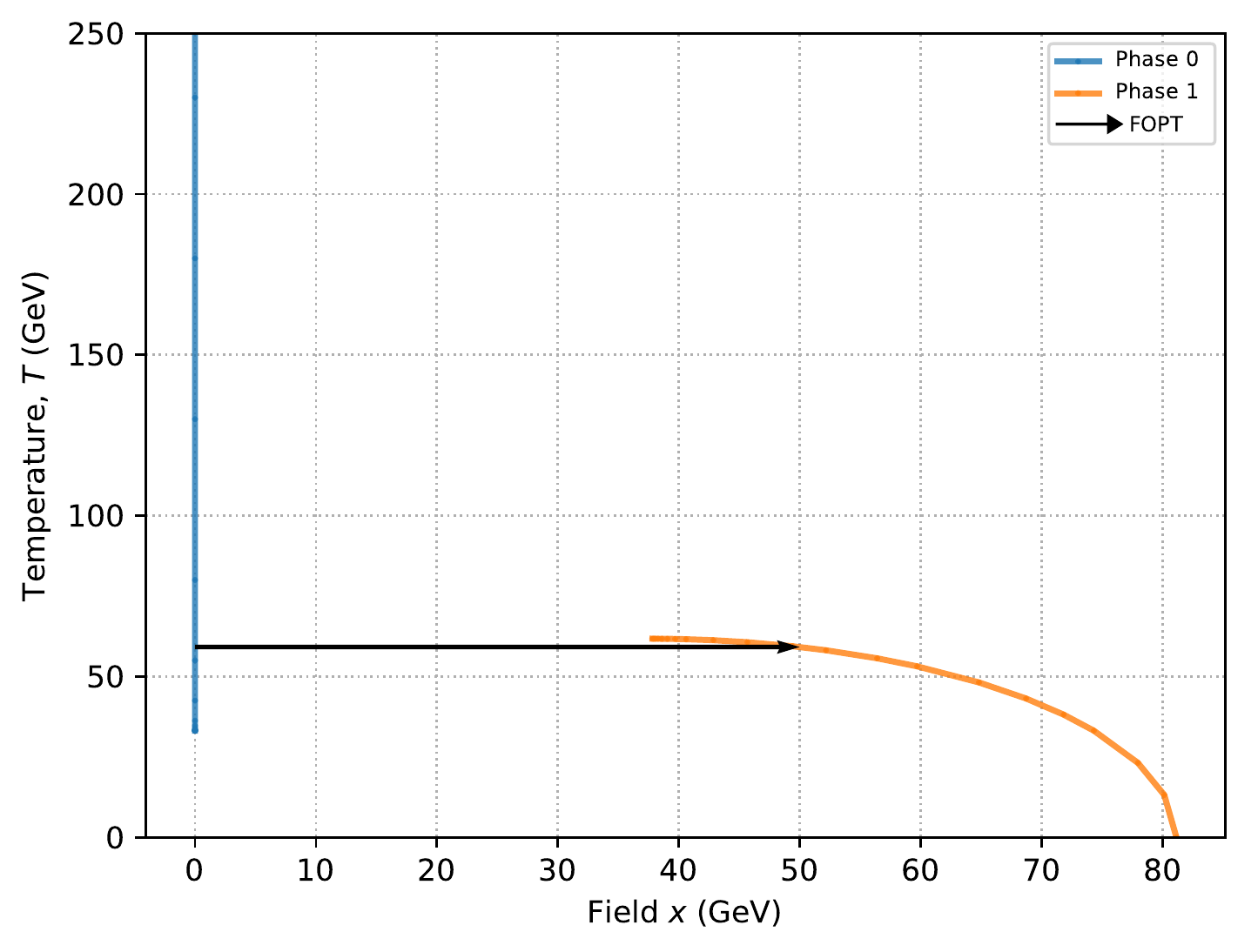}
\caption{The one-dimensional potential at different temperatures (left) and the subsequent phase structure of the model (right). On the right panel, the lines show the field values at a particular minimum as a function of temperature. The arrows indicate that at that temperature the two phases linked by the arrows are degenerate and thus that a FOPT could occur in the direction of the arrow. The label $x$ was autogenerated by our plotting code. In this example it stands for $\phi$.}
\label{fig:1d}
\end{figure}

To exhibit the usage of \pt, we consider a one-dimensional potential,
\begin{equation}
    V(\phi, T) = (c T^2 - m^2) \phi^2 + \kappa \phi^3 + \lambda \phi^4
\end{equation}
for $\phi \ge 0$ and $\kappa < 0$ as a simple dummy model, without physical meaning. We consider the specific numerical constants,
\begin{equation}
    V(\phi,T) = (0.1 T^2 - 100)\phi^2  - 10\phi^3 +0.1\phi^4.
\end{equation}
The simplicity of the model means that it can be solved analytically, giving
\begin{equation}
T_C = \sqrt{\frac{\kappa^2 + 4 \lambda m^2}{4c\lambda}} = 59.1608
\end{equation}
with minima at $0$ and $-\kappa / (2\lambda) = 50$ separated by a barrier at $-\kappa / (4\lambda) = 25$. We show the potential and the phase transition in \figref{fig:1d}. We find a numerical result of $T_C = 59.1608$ with minima at $4.92767\mathrm{e}{-6}$ and $50.0003$, i.e.~the numerical result  matches the exact analytic result up to at least six significant figures implying excellent agreement.\footnote{The agreement on $T_C$ is often much better than the nominal relative precision, \code{TC_rel_tol = 1e-4}, as the \code{toms748_solve} root-finding algorithm converges rapidly in many cases. However in principle it can be worse than the nominal precision, because there is  uncertainty on the two minima.} This result may be reproduced by
\begin{lstlisting}[language=bash]
$ ./bin/run_1D_test_model -d
\end{lstlisting}
The argument --- \code{-d} --- indicates that we want to produce debugging information including plots named \code{*1D_test_model.pdf}.

\subsection{Two-dimensional test model}\label{sec:2D_test_model}

To compare the performance of \pt with \cosmo, we adopt the test model with two scalar fields in \cosmo. The tree-level potential is
\begin{equation}
    V(\phi_1,\phi_2) = \frac{1}{8}\frac{m_1^2}{v^2}(\phi_1^2-v^2)^2 +\frac{1}{8}\frac{m_2^2}{v^2}(\phi_2^2-v^2)^2 - \mu^2 \phi_1\phi_2.
\end{equation}
The resulting field-dependent mass matrix for the pair of scalar fields is\footnote{We correct a typo for the off-diagonal elements in \refcite{Wainwright:CosmoTransition}. The typo was not present in the \cosmoversion source code.}
\begin{equation}
    M^2 (\phi_1,\phi_2) =
    	\begin{pmatrix}
		 \frac{m_1^2}{2v^2} (3\phi_1^2-v^2) & -\mu^2 \\
		 -\mu^2 & \frac{m_2^2}{2v^2} (3\phi_2^2-v^2) \\
	\end{pmatrix}.
\end{equation}
Here $m_1$, $m_2$, $v$ and $\mu$ take the values listed in Table 1 of \refcite{Wainwright:CosmoTransition}, i.e.\ $120\gev$, $50 \gev$, $246 \gev$ and $25 \gev$, respectively. In addition to the scalar fields in the tree-level potential, we add a scalar boson $X$ with $n_X=30$ degrees of freedom and the field-dependent mass
\begin{equation}
    m_X(\phi_1,\phi_2) = y_A^2(\phi_1^2+\phi_2^2)+y_b^2\phi_1\phi_2,
\end{equation}
where $Y_A^2=0.1$ and $Y_B^2=0.15$. We set the renormalization scale in the one-loop potential to $246 \gev$.

We compare our results against \cosmoversion (at present the latest version) using its default settings.\footnote{Note, however, that numerical results for this potential with \cosmoversion are not consistent with results from previous versions of \cosmo documented in \refcite{Wainwright:CosmoTransition}.} After checking our implementation of the potential, by making sure we obtain exactly the same potential for given field values and temperature, we test the critical temperature and true and false vacuums found by our code. Our results may be reproduced by
\begin{lstlisting}[language=bash]
$ ./bin/run_2D_test_model -d
\end{lstlisting}
The argument --- \code{-d} --- indicates that we want to produce debugging information including plots named \code{*2D_test_model.pdf}.

\begin{table}[h]
  \centering
  \begin{tabular}{ccccccc}
    \toprule
    				        & $T_C$ & False VEV         & True VEV		        & Time (seconds)\\
    \midrule
    \cosmo                  & 109.4 & (220.0, $-150.0$) & (263.5, 314.7)        & 3.51\\
    \midrule
    \multirow{2}{*}{\pt} 	& \multirow{2}{*}{109.4} & (220.0, $-150.0$)	& (263.5, 314.7)        & \multirow{2}{*}{0.04}\\ 
          	                &                        & (220.0, $-150.0$)	& ($-263.5$, $-314.7$)  &  \\
    \bottomrule
  \end{tabular}
  \caption{Results and elapsed time of \pt and \cosmo for the 2D test model. All dimensionful quantities are in GeV. The base of the VEV is $(\phi_1,\phi_2)$.}
  \label{tab:2d}
\end{table}

\begin{figure}[t]
\centering
\includegraphics[height=.34\textwidth]{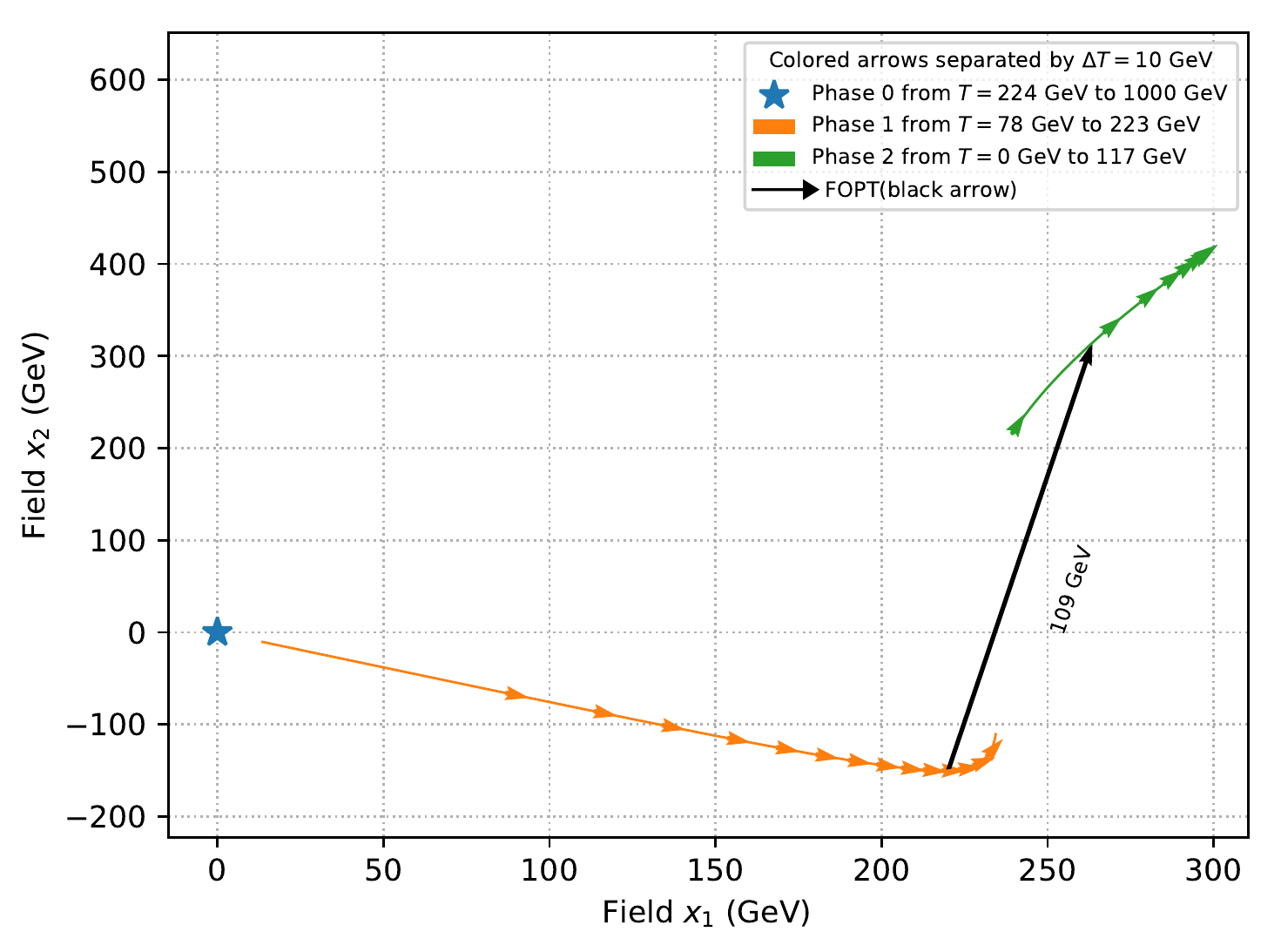}
\includegraphics[height=.34\textwidth]{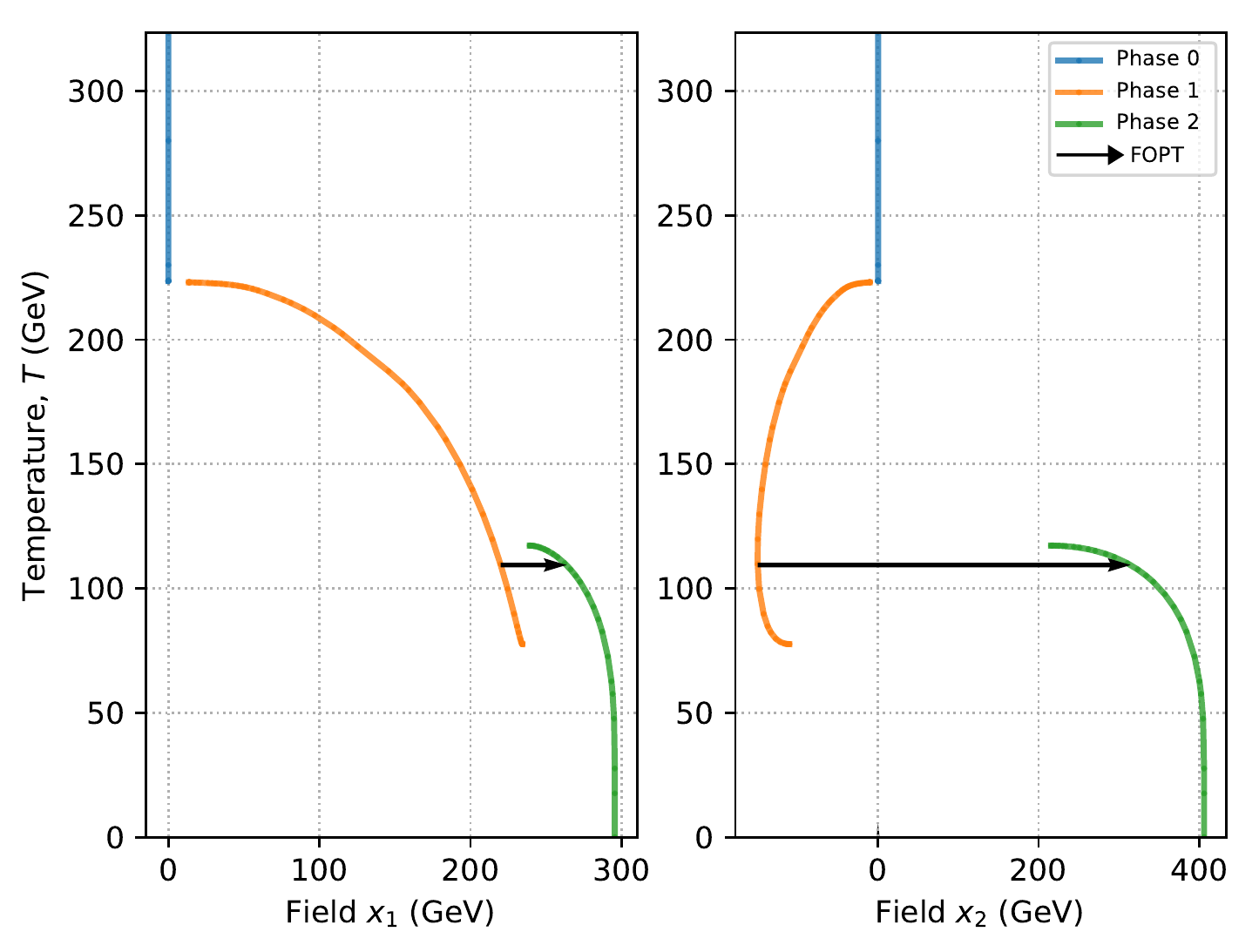}
\caption{Phase structure of the 2D test model. In the left panel, the changes in field values at a particular minimum with decreasing temperature are indicated by colored arrows in steps of $\Delta T = 10 \gev$. The star represents a phase that does not change with temperature. The meaning of the right panel is same as the right panel in \figref{fig:1d}. The labels $x_1$ and $x_2$ were autogenerated by our plotting code. In this example they stand for $\phi_1$ and $\phi_2$, respectively.}
\label{fig:2d}
\end{figure}

We show in \tabref{tab:2d} that our results agree extremely well\footnote{In fact in this case they agree up to 9 significant figures, with default settings.}  with \cosmo for the critical temperature and field values for the first possible transition; however, our code is about 100 times faster for this problem.\footnote{%
We computed timings of \pt in \code{C++} using the average of 1000 repeats in \code{example/time.cpp} and \code{example/time.hpp} with a desktop with an Intel Core i7-6700 CPU @ 3.40GHz processor. For \cosmo, we averaged the run time of \code{"for i in range(1000): model1().calcTcTrans()"} inside \code{examples/testModel1.py} of \cosmoversion on the same machine.
}
The phase transitions in this problem are shown in \figref{fig:2d}. We see in the left panel a parametric plot of the two fields as functions of temperature. The distinct phases are shown by blue, orange and green colors, and the FOPT is shown by a black arrow. In the right panel, we see each field as a function of temperature.

Note, however, that \pt found two transitions whereas \cosmo only found one. Due to the discrete symmetry, $\phi \to -\phi$, there is in fact a cousin transition $(220.0, -150.0)\to(-263.5, -314.7)$ that accopanies the one found by \cosmo (see \secref{sec:symmetries}). We always save all the transitions in the results, as we do not know a priori which transition has the greatest tunnelling probability. For this example, we calculated the bounce actions at $T=100\gev$ using \cosmo to be $S_E=102659.3$ for $(224.5, -148.3)\to(275.3, 351.0)$ and $S_E=1402952.5$ for $(224.5, -148.3)\to(-275.3, -351.0)$.  In the \cosmo example, the duplicated phase and thus the cousin transition is removed by forbidding $\phi_1<0$ and it returns only the transition with the smaller action.\footnote{In the code, it is $\phi_1<-5$ to allow for slight inaccuracy in the location of the phase.} However, this appears to be a coincidence: forbidding $\phi_2 < 0$ instead of $\phi_1 < 0$ would result in \cosmo finding only the transition with the larger action\footnote{In fact, if $\phi_2 < 0$ is forbidden, \cosmoversion would only find Phase 2 shown in \figref{fig:2d}.}. Note that in the automatically generated plots, such as \figref{fig:2d}, the symmetric cousin transitions are not shown, as their inclusion makes the plots hard to read.

\subsection{\ztwo Scalar Singlet Model}
\label{sec:ssm}

As a more realistic two-dimensional example, we consider the \ztwo scalar singlet extension of the SM. Unlike the SM, this model can accommodate a $125\gev$ Higgs and a first order phase transition, as well as possibly providing a viable dark matter candidate. The potential consists of the ordinary SM Higgs potential and gauge- and \ztwo invariant interactions involving a real singlet,
\begin{equation}
V(H, s) = \mu^2_H |H|^2 + \lambda_H |H|^4 + \lambda_{Hs} |H|^2 s^2 + \tfrac12 \mu^2_s s^2 + \tfrac14 \lambda_s s^4.
\end{equation}
After EWSB, the real scalar obtains a tree-level mass $m_S^2 = \mu^2_s + \tfrac12 \lambda_{Hs} v^2$. We implement one-loop and daisy corrections to this tree-level potential. Under certain circumstances, a first order transition between an EW preserving and an EW breaking vacuum may be possible and the critical temperature can be found analytically; see, e.g., \refcite{Vaskonen:2016yiu} for an analysis of the PTs in this model.

\begin{figure}[t]
\centering
\includegraphics[height=.34\textwidth]{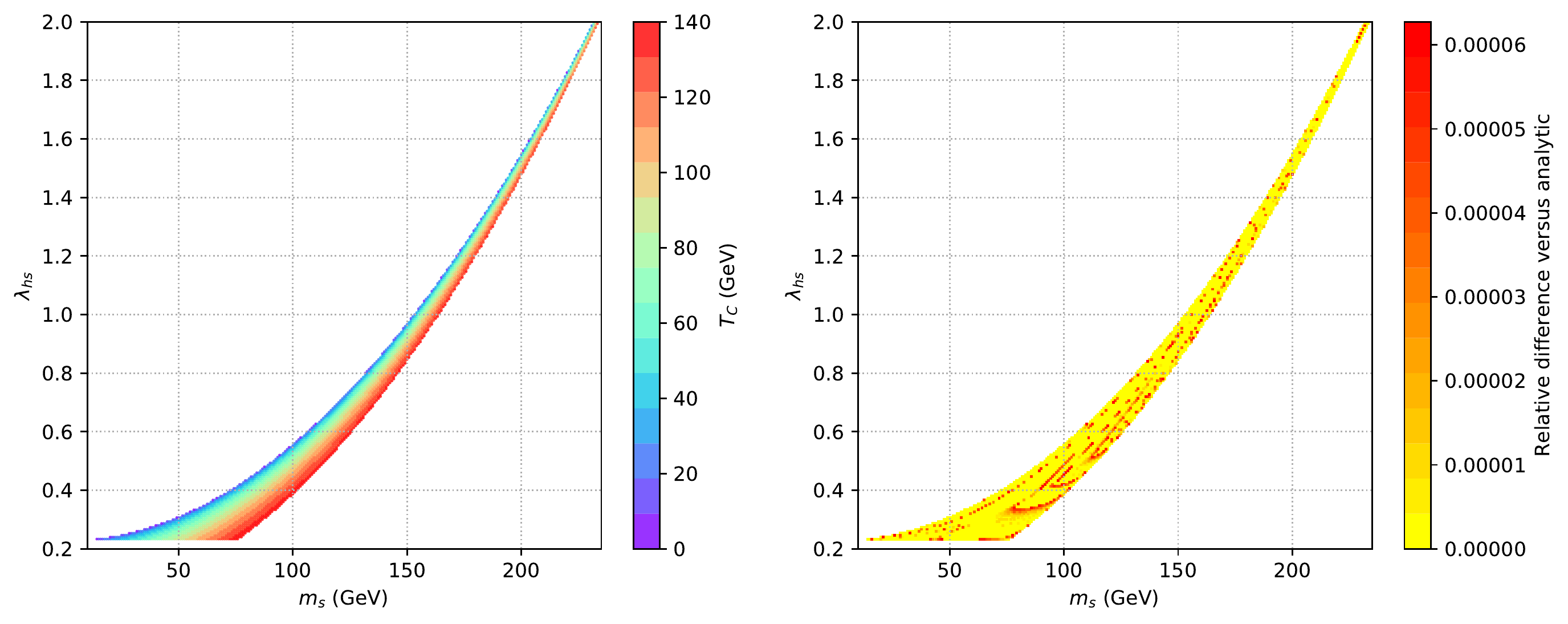}
\caption{The critical temperature, $T_C$, as a function of $\lambda_{hs}$ and $m_s$ in the \ztwo SSM calculated by \pt (left) and the relative difference versus the analytic result (right). There is no FOPT in the white regions.}
\label{fig:ssm}
\end{figure}

We reproduce the behaviour of the critical temperature as a function of $\lambda_{Hs}$ and $m_S$ that was found in \refcite{Vaskonen:2016yiu} (see, e.g., \figurename~1 of \refcite{Vaskonen:2016yiu}) in \figref{fig:ssm}. On the right panel, we show that the differences between numerical results from \pt and the analytic formulae are at most about $0.01\%$. To reproduce it,
\begin{lstlisting}[language=bash]
$ ./bin/scan_Z2_scalar_singlet_model
$ python make_plots/compare_against_analytic_z2.py
\end{lstlisting}
The first command scans the relevant parameter space and writes a data file named \code{Z2ScalarSingletModel_Results.txt} that the second command plots in \code{ms_lambda_Z2_SSM.pdf}.

\subsection{Two-Higgs-Doublet Models} \label{sec:2HDM}

We next consider three Two-Higgs-Doublet Models (2HDMs) which are implemented in \bsmpt. We do not reimplement their potentials; instead, we link \pt directly to the models implemented in \bsmpt and directly call the \bsmpt potentials. The general 2HDM tree-level scalar potential is
\begin{equation}
\begin{split}
V  ={}& m_{11}^2 |H_1|^2 + m_{22}^2 |H_2|^2
     - m_{12}^2\left(H_1^\dagger H_2 + \text{h.c.}\right)
    + \tfrac12 \lambda_1 |H_1|^4
    + \tfrac12 \lambda_2 |H_2|^4\\
    & + \lambda_3 |H_1|^2 |H_2|^2
    + \lambda_4 |H_1^\dagger H_2|^2
    + \tfrac12 \lambda_5 \left[\left(H_1^\dagger H_2\right)^2 + \text{h.c.}\right].
\end{split}
\end{equation}
The two Higgs doublets take the form
\begin{equation}
\begin{split}
H_1 &= \frac{1}{\sqrt{2}}
         \begin{pmatrix}
		   \rho_1 + i\eta_1 \\
		   v_1 + \zeta_1 + i\psi_1 \\
		  \end{pmatrix} \\
H_2 &= \frac{1}{\sqrt{2}}
         \begin{pmatrix}
		   v_{\rm CB} + \rho_2 + i\eta_2 \\
		   v_2 + \zeta_2 + i(v_{\rm CP}+\psi_2) \\
		  \end{pmatrix} \\
\end{split}
\end{equation}
where $\rho_i$ and $\eta_i$ $(i=1,2)$ represent the charged field components, $\zeta_i$ and $\psi_i$ $(i=1,2)$ indicate the neutral CP-even and CP-odd fields, and the VEVs $v_i$ $(i=1,2,{\rm CB},{\rm CP})$ are real.

First, we consider the real, CP-conserving 2HDM (R2HDM) with the parameters
\begin{equation}
\begin{array}{lcllcl}
\mbox{Type} & = & 1 & \; \tan\beta & = & 4.63286 \\
\lambda_1 & = & 2.740594787 & \; \lambda_2 & = & 0.2423556498  \\
\lambda_3 & = & 5.534491052 & \; \lambda_4 & = & -2.585467181  \\
\lambda_5 & = & -2.225991025 & \; m_{12}^2  & = & 7738.56 \gev^2.
\end{array}  \label{eq:c2hdmex}
\end{equation}

Second, we consider the complex, CP-violating 2HDM (C2HDM) with the parameters
\begin{equation}
\begin{array}{lcllcl}
\mbox{Type} & = & 1 & \; \tan\beta & = & 4.64487 \\
 \lambda_1 & = & 3.29771 & \;
\lambda_2 & = & 0.274365  \\
\lambda_3 & = & 4.71019 & \; \lambda_4 & = & -2.23056
\\
\mbox{Re} (\lambda_5) & = & -2.43487 & \; \mbox{Im} (\lambda_5) & = & 0.124948
  \\
\mbox{Re} (m_{12}^2) & = & 2706.86 \gev^2 \;.
\end{array}  \label{eq:c2hdmex}
\end{equation}
Note that in this case we consider a complex phase for $\lambda_5$.
Lastly, we consider the Next-to-2HDM (N2HDM) with the parameters
\begin{equation}
\begin{array}{lcllcl}
\mbox{Type} & = & 1 & \; \tan\beta & = & 5.91129 \\
\lambda_1 & = & 0.300812  & \; \lambda_2 & = & 0.321809  \\
\lambda_3 & = & -0.133425 & \; \lambda_4 & = & 4.11105\\
\lambda_5 & = & -3.84178  & \; \lambda_6 & = & 9.46329\\
\lambda_7 & = & -0.750455 & \; \lambda_8 & = & 0.743982\\
v_s       & = & 293.035 \gev & \; m_{12}^2  & = & 4842.28 \gev^2.
\end{array}  \label{eq:c2hdmex}
\end{equation}
Compared to the 2HDM models, this model contains an extra complex singlet with \ztwo symmetric interactions,
\begin{equation}
\tfrac12 m_S^2 S^2 + \tfrac18 \lambda_6 S^4 +
\tfrac12 \lambda_7 |H_1|^2 S^2 +
\tfrac12 \lambda_8|H_2|^2  S^2.
\end{equation}
The singlet field $S$ is expanded as $S=v_S+\zeta_S$ with $v_S$ denoting singlet VEV.

To run our code on these models,
\begin{lstlisting}[language=bash]
$ ./bin/run_R2HDM
$ ./bin/run_C2HDM
$ ./bin/run_N2HDM
\end{lstlisting}
Note that these examples require \bsmpt; see \secref{sec:download_etc} for installation instructions. Regarding finding the critical temperature, our code is slower than \bsmpt as we perform more checks and thoroughly map out the whole phase structure.

\begin{table}[h]
  \centering
  \begin{tabular}{ccccccc}
    \toprule
    				     & $T_C$    & False VEV          & True VEV		   \\
    \midrule
    R2HDM\\
    \midrule
    \bsmpt               & 155.283  & - 	& (0, $-54.2$, $-182.9$, 0)\\
    \cmidrule(r){2-4}
    \multirow{3}{*}{\pt} & 155.284  & (0, 0, 0, 0)	& (0, 54.2, 182.9, 0) \\
                         & \multirow{2}{*}{161.205} & (0, 45.9, 128.7, 0) & (0, 50.5, 152.6, 0) \\
                         &                          & (0, 45.9, 128.7, 0) & (0, $-50.5$, $-152.6$, 0) \\
    \midrule
    C2HDM\\
    \midrule
    \bsmpt     & 145.569  & - 	& (0, 49.9, 194.5, $-1.3$)\\
    \pt        & 145.575  & (0, 0, 0, 0) & (0, 49.9, 194.5, $-1.3$) \\
    \midrule
    N2HDM\\
    \midrule
    \bsmpt  & 121.06  & - 	              & (0, 0, $-32.5$, $-176.3$, $-297.1$)\\
    \cmidrule(r){2-4}
    \multirow{2}{*}{\pt} & \multirow{2}{*}{120.73}  & (0, 0, 0, 0, 301.0) & (0, 0, 32.7, 177.6, 297.0) \\
                         &                          & (0, 0, 0, 0, 301.0) & (0, 0, 32.7, 177.6, $-297.0$) \\
    \bottomrule
  \end{tabular}
  \caption{Results of \pt and \bsmptversion for the R2HDM, C2HDM and N2HDM benchmark points. All dimensionful quantities are in GeV. The basis of the VEVs is $(v_{\rm CB},v_1,v_2,v_{\rm CP})$ for R2HDM and C2HDM, and $(v_{\rm CB},v_{\rm CP},v_1,v_2,v_S)$ for N2HDM. The dash for the false VEV from \bsmpt indicates that it doesn't calculate the false VEV (it assumes that it lies at the origin).}
\label{tab:2hdm}
\end{table}

We show results from all three models with our code and \bsmpt in \tabref{tab:2hdm}. In \bsmpt, the tolerance of the bisection method used to locate the $T_C$ is $0.01\gev$ , while our default relative precision \code{TC_rel_tol} is 0.01\%. Thus our results for R2HDM and C2HDM benchmark points are in agreement with that from \bsmpt in the range of these errors. In addition, we find an extra group of transitions at $T_C=161.205\gev$ for R2HDM, which is missed by \bsmpt because it focuses only on the transition that starts in the symmetric vacuum. For the N2HDM benchmark point, the discrepancy of $T_C$ is larger than the expected precision. This is related to the fact that \bsmpt assumes that the false vacuum lies at the origin and with its default settings \bsmpt misses the minimum around at $(0, 0, 0, 0, 301.0)$ at $T\simeq 121.00\gev$, mistakenly treating the minimum around at $(0, 0, 32.5, 176.5, 297.1)$ as the deepest minima at that temperature.\footnote{The results from \pt and \bsmpt are all sensitive to the random number generation.} By increasing the number of random starting points for finding multiple local minima in \bsmpt, it finds this minima and agrees with our result, $T_C=120.73\gev$, though the agreement may be partially an accident since the false vacuum does not lie at the origin, contrary to the assumption in \bsmpt.

\subsection{Next-to-Minimal Supersymmetric Standard Model}
\label{sec:nmssm}

\begin{figure}[t]
\centering
\begin{minipage}[c]{\textwidth}
    \raggedright
    \includegraphics[height=.34\textwidth]{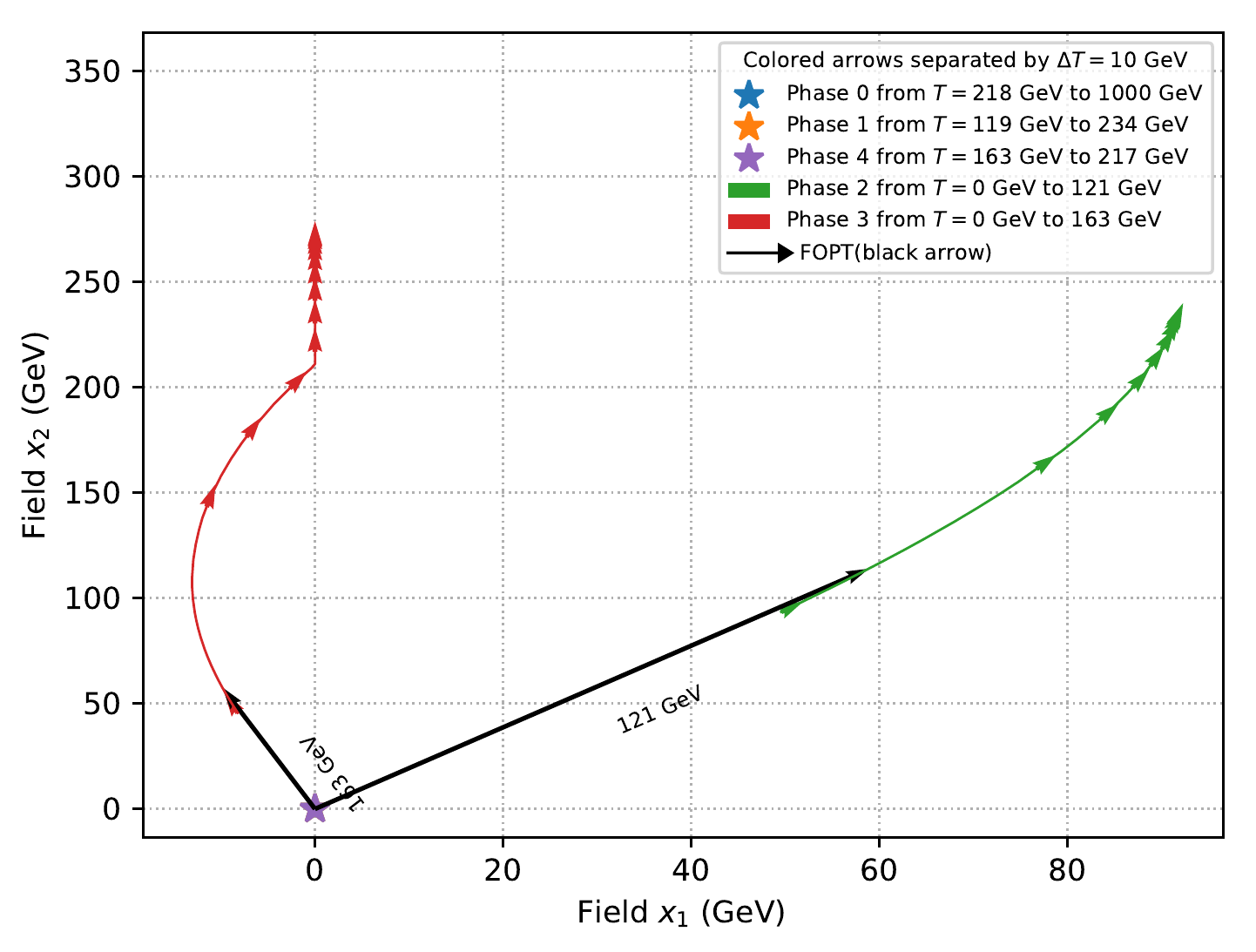}
    \includegraphics[height=.34\textwidth]{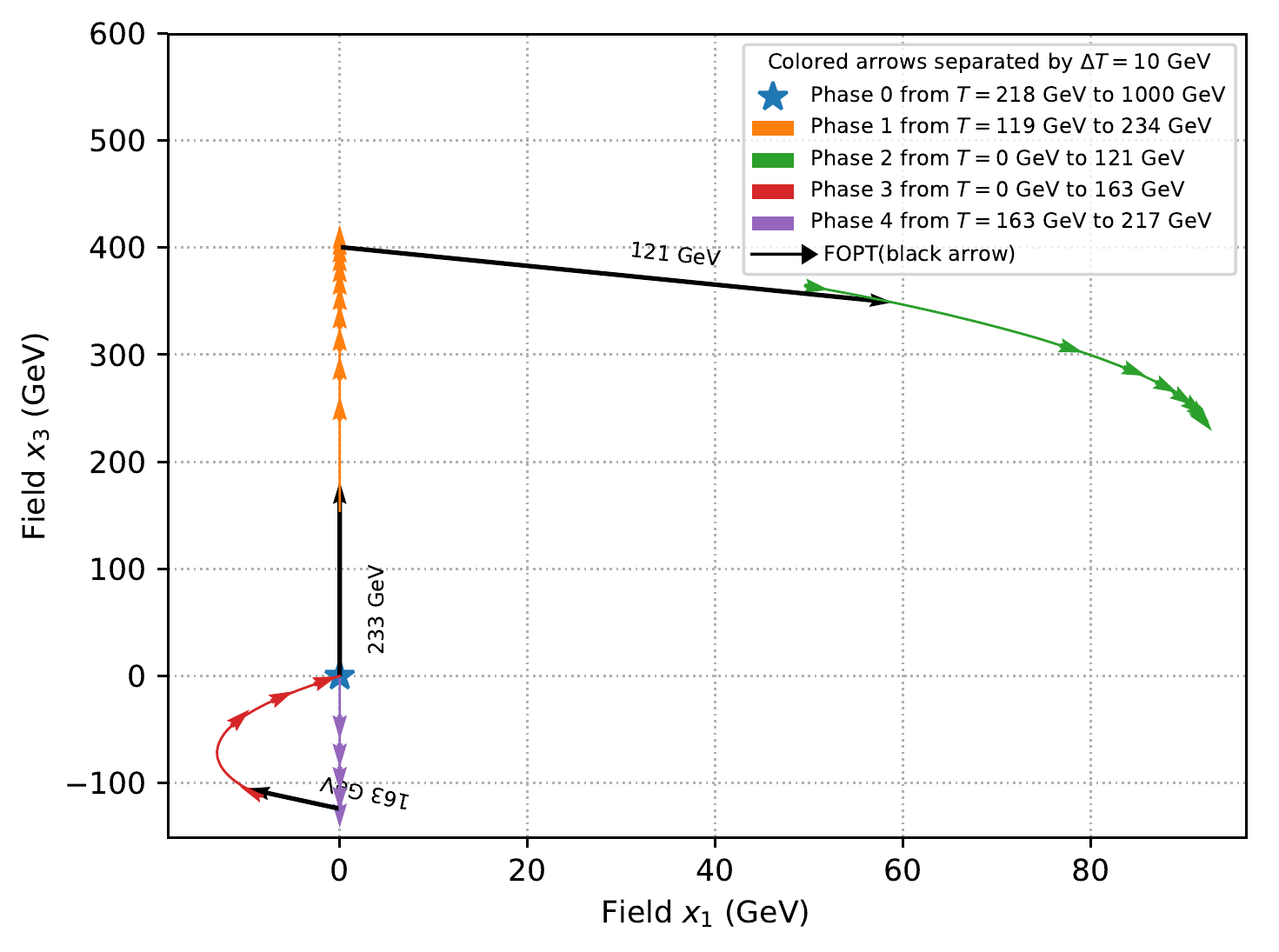}\\
    \includegraphics[height=.34\textwidth]{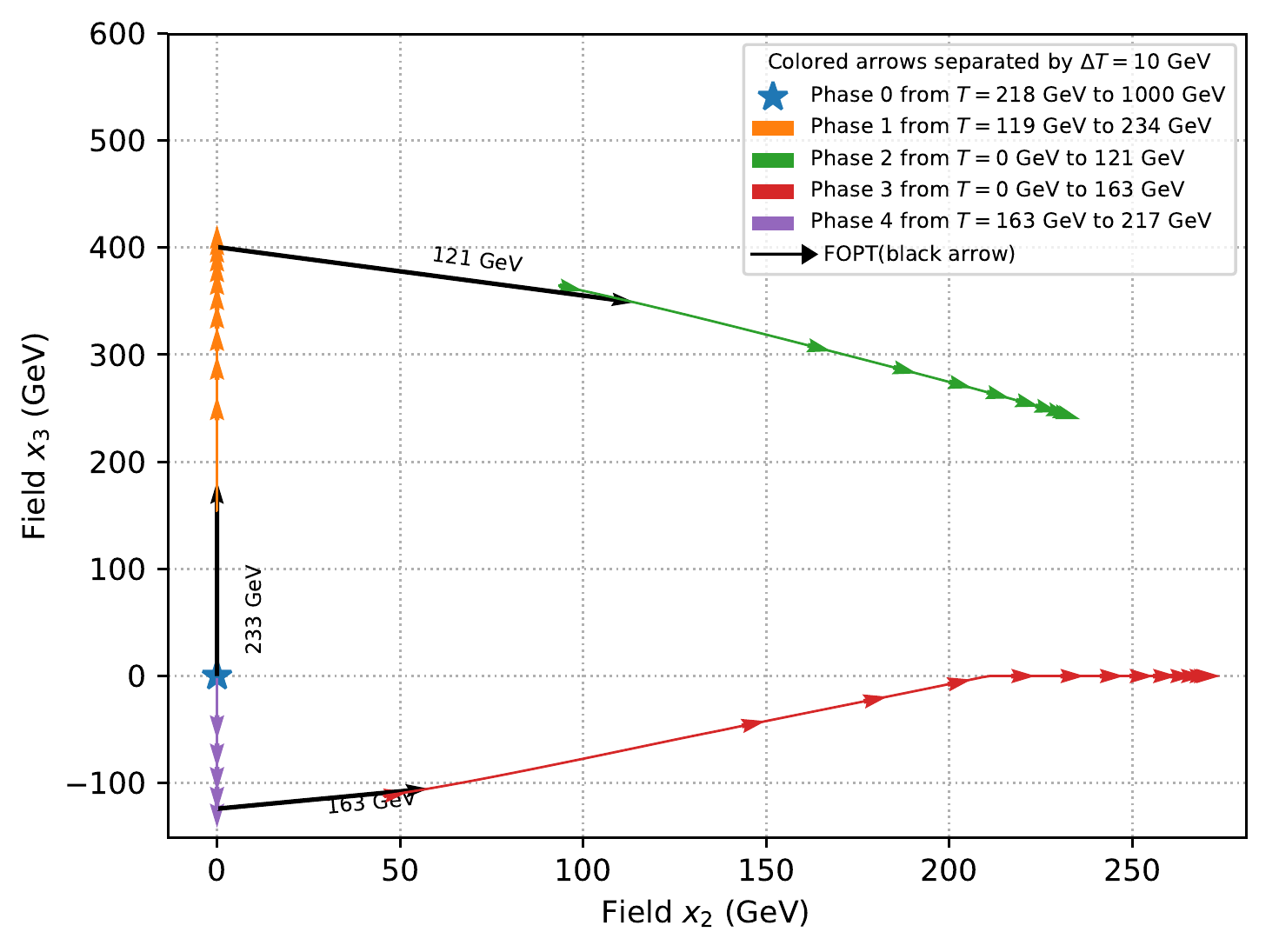}
    \includegraphics[height=.34\textwidth]{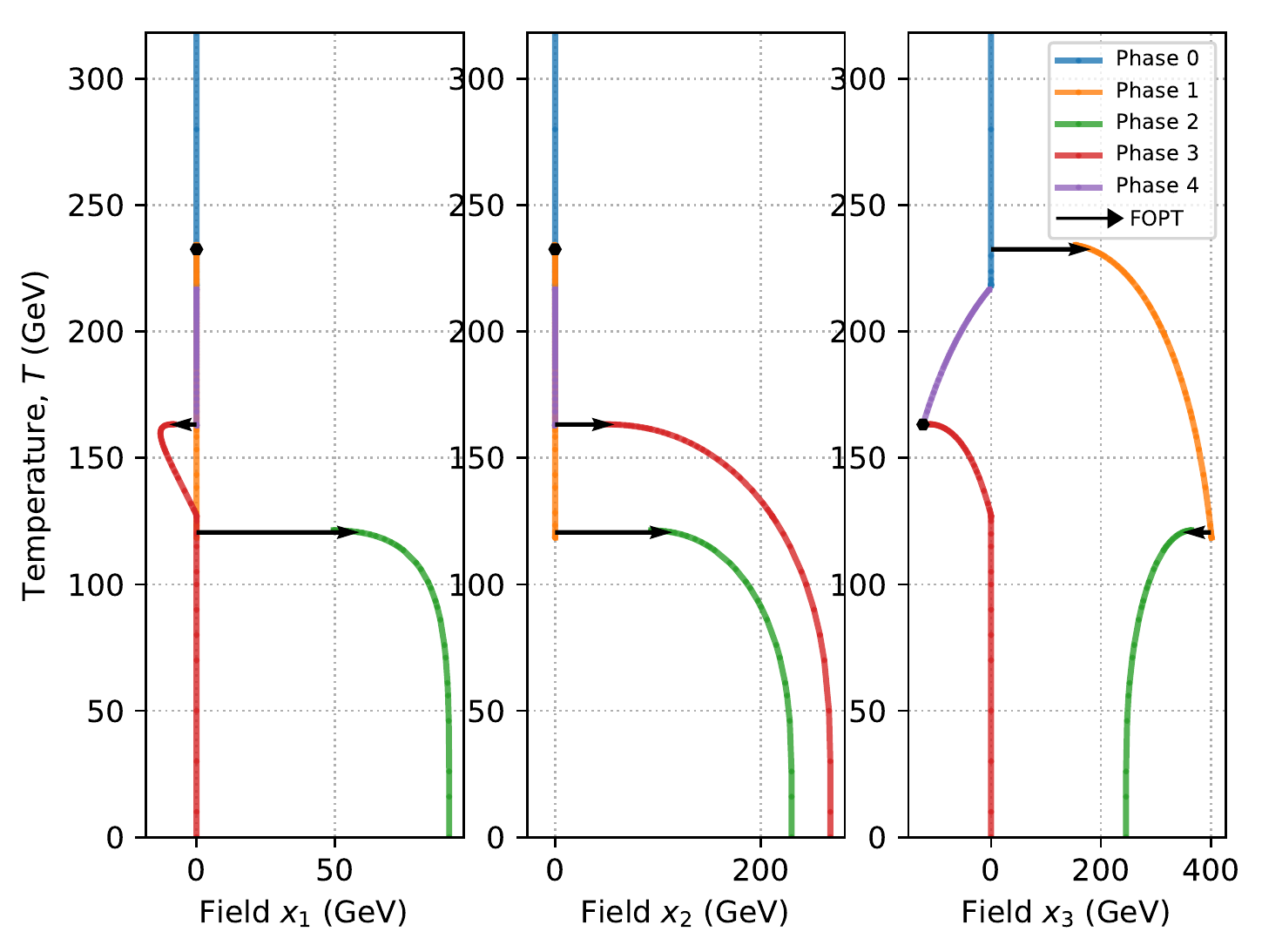}
\end{minipage}
\caption{Phase structure of the NMSSM. The meanings of the upper panels and the lower left panel is same to the left panel of \figref{fig:2d}. The meanings of the right panel is similar to the right panel of \figref{fig:1d}. The dots represent transitions that do not change the corresponding field values. The labels $x_1$, $x_2$ and $x_3$ were autogenerated by our plotting code. In this example they stand for $h_u$, $h_d$ and $s$, respectively.}
\label{fig:nmssm}
\end{figure}

In \refcite{Athron:2019teq} we explored the Next-to-Minimal Supersymmetric Standard Model (NMSSM) with a preliminary version of \pt. We include our NMSSM model (in which we match the NMSSM to a model with two-Higgs doublets and a singlet; see  \refcite{Athron:2019teq} for further details) as an example. The effective potential depends upon \fs (see \secref{sec:download_etc} for the relevant installation instructions) for the matching conditions, tadpole equations and field-dependent masses. We show a benchmark point from this model in \figref{fig:nmssm}. To reproduce it,
\begin{lstlisting}[language=bash]
$ ./bin/run_THDMIISNMSSMBCsimple
\end{lstlisting}
which will solve this problem and produce the figures shown in \figref{fig:nmssm} in files named \code{*THDMIISNMSSMBCsimple.pdf}.

Because this model point is particularly pathological, there is a chance that \pt may miss Phase 4 and part of Phase 3 shown in \figref{fig:nmssm}. This is because at $T=127.1\gev$, there exists a saddle point at $(h_u=0\gev, h_d=211.0\gev, s=0\gev)$ near a minimum. When \pt traces the phases near this saddle point, it may correctly find the real minimum near the saddle point or mistake the saddle point for a minimum.  In the latter case, when \pt discovers that the Hessian isn't positive semi-definite, it stops
tracing Phase 3 at the saddle point, and then will miss Phase 4. We will address this problem by improving the \code{PhaseFinder::find_min} method in a future version so that it cannot mistake saddle points for minima. The behaviour of \pt in these cases isn't deterministic since it depends on random number generation for locating minima.

\section{Conclusions}
\label{sec:conclusions}
In this paper we have presented \pt, a fast and reliable \code{C++}
software package for finding the cosmological phases and the critical
temperatures for phase transitions. For any user supplied potential,
\pt first maps out all of the phases over the relevant temperature
range\footnote{The upper and lower bound on the temperature is chosen
  by the user, though by default these are taken to be $T=0$ and $T=1$
  TeV, which are the values relevant for electroweak phase
  transitions.}. Once the phases have been identified \pt checks each
pair of phases to see if there may be a first order phase transition between them and calculates the critical temperature.

To do this \pt uses a similar algorithm to that of \cosmo, but our
implementation has a number of advantages: i) \pt is orders of
magnitude faster, as we have demonstrated in
\secref{sec:2D_test_model}; ii) \pt has been carefully designed to
provide clear error information for cases where a solution cannot be
found and has many adjustable settings which may be varied to find
solutions in such cases; iii) \pt will be maintained by an active team
of developers (the authors of this manual) and is distributed with a
suite of unit tests to validate the code and ensure
results do not change unintentionally as the code evolves. In addition
\pt has been designed so that it can be easily linked with \fs
spectrum generators, making it easy to embed it in a much wider
tool set for investigating the phenomenology of an arbitrary BSM
extension.  Furthermore we have also made it possible to link to
potentials implemented in \bsmpt, another tool for finding critical
temperatures.  This makes it possible for users to compare results in
both \bsmpt and \pt, as we have demonstrated for two Higgs doublet and
two Higgs plus singlet extensions of the Standard Model in
\secref{sec:2HDM}. This is particularly useful since \bsmpt has a
complementary approach, implementing a method that is simpler and
faster, but finds exactly one transition that starts from the
symmetric vacuum. Therefore \bsmpt will miss other transitions and the
transition it does find may not be in the cosmological history, while
\pt will find all potential transitions.

With \pt it is therefore very easy to find the phases and critical
temperatures of arbitrary Standard Model extensions.  This is an
important and useful tool for investigating both electroweak
baryogenesis and gravitational waves, and can be used in thorough
phenomenological investigations of realistic Standard Model extensions
with many parameters, as we have demonstrated in
Ref.\ \cite{Athron:2019teq} where an early prototype of \pt was used.
Furthermore \pt is part of a wider goal to develop a set of tools that
can be used to automatically calculate the baryon asymmetry of the
Universe, predict the stochastic gravitational wave background from
first order phase transitions and test this against observations from
gravitational wave experiments.

\section*{Acknowledgments}
We thank Graham White for advice and expertise passed on in other phase transition projects, Michael Bardsley for restructuring of our \ep library so that it will be easier in future to use this with \pt, and Giancarlo Pozzo for extensive early testing of the code.  We also thank both Michael and Graham for crucial  participation in our the wider goal of a complete package/set of codes for studying EWBG and GWs. The work of P.A.~is supported by the Australian Research Council Future Fellowship grant FT160100274.  P.A.~also acknowledges the hospitality of Nanjing Normal University while working on this manuscript.  The work of C.B. and Y.Z. was supported by the Australian Research Council through the ARC Centre of Excellence for Particle Physics at the Tera-scale CE110001104.  The work of P.A. and C.B. is also supported with the Australian Research Council Discovery Project grant DP180102209.

\clearpage

\addcontentsline{toc}{section}{References}

\bibliographystyle{JHEP}
\bibliography{bibliography}

\providecommand{\href}[2]{#2}\begingroup\raggedright\begin{thebibliography}{10}

\bibitem{Mazumdar:2018dfl}
A.~Mazumdar and G.~White, \emph{{Review of cosmic phase transitions: their
  significance and experimental signatures}},
  \href{https://doi.org/10.1088/1361-6633/ab1f55}{\emph{Rept. Prog. Phys.}
  {\bfseries 82} (2019) 076901},
  [\href{https://arxiv.org/abs/1811.01948}{{\ttfamily 1811.01948}}].

\bibitem{Kirzhnits:1972iw}
D.~A. Kirzhnits, \emph{{Weinberg model in the hot universe}}, {\emph{JETP
  Lett.} {\bfseries 15} (1972) 529--531}.

\bibitem{Kirzhnits:1972ut}
D.~A. Kirzhnits and A.~D. Linde, \emph{{Macroscopic Consequences of the
  Weinberg Model}},
  \href{https://doi.org/10.1016/0370-2693(72)90109-8}{\emph{Phys. Lett.}
  {\bfseries 42B} (1972) 471--474}.

\bibitem{Pati:1974yy}
J.~C. Pati and A.~Salam, \emph{{Lepton Number as the Fourth Color}},
  \href{https://doi.org/10.1103/PhysRevD.10.275,
  10.1103/PhysRevD.11.703.2}{\emph{Phys. Rev.} {\bfseries D10} (1974)
  275--289}.

\bibitem{Fritzsch:1974nn}
H.~Fritzsch and P.~Minkowski, \emph{{Unified Interactions of Leptons and
  Hadrons}}, \href{https://doi.org/10.1016/0003-4916(75)90211-0}{\emph{Annals
  Phys.} {\bfseries 93} (1975) 193--266}.

\bibitem{Georgi:1974my}
H.~Georgi, \emph{{The State of the Art—Gauge Theories}},
  \href{https://doi.org/10.1063/1.2947450}{\emph{AIP Conf. Proc.} {\bfseries
  23} (1975) 575--582}.

\bibitem{Gursey:1975ki}
F.~Gursey, P.~Ramond and P.~Sikivie, \emph{{A Universal Gauge Theory Model
  Based on E6}},
  \href{https://doi.org/10.1016/0370-2693(76)90417-2}{\emph{Phys. Lett.}
  {\bfseries 60B} (1976) 177--180}.

\bibitem{Georgi:1974sy}
H.~Georgi and S.~L. Glashow, \emph{{Unity of All Elementary Particle Forces}},
  \href{https://doi.org/10.1103/PhysRevLett.32.438}{\emph{Phys. Rev. Lett.}
  {\bfseries 32} (1974) 438--441}.

\bibitem{Cvetic:1996mf}
M.~Cvetic and P.~Langacker, \emph{{New gauge bosons from string models}},
  \href{https://doi.org/10.1142/S0217732396001260}{\emph{Mod. Phys. Lett.}
  {\bfseries A11} (1996) 1247--1262},
  [\href{https://arxiv.org/abs/hep-ph/9602424}{{\ttfamily hep-ph/9602424}}].

\bibitem{Cvetic:1997wu}
M.~Cvetic and P.~Langacker, \emph{{Z' Physics and Supersymmetry}},
  \href{https://doi.org/10.1142/9789812839657_0012}{\emph{Adv. Ser. Direct.
  High Energy Phys.} {\bfseries 18} (1998) 312--331},
  [\href{https://arxiv.org/abs/hep-ph/9707451}{{\ttfamily hep-ph/9707451}}].

\bibitem{Cleaver:1998gc}
G.~Cleaver, M.~Cvetic, J.~R. Espinosa, L.~L. Everett, P.~Langacker and J.~Wang,
  \emph{{Physics implications of flat directions in free fermionic superstring
  models 1. Mass spectrum and couplings}},
  \href{https://doi.org/10.1103/PhysRevD.59.055005}{\emph{Phys. Rev.}
  {\bfseries D59} (1999) 055005},
  [\href{https://arxiv.org/abs/hep-ph/9807479}{{\ttfamily hep-ph/9807479}}].

\bibitem{Cleaver:1998sm}
G.~Cleaver, M.~Cvetic, J.~R. Espinosa, L.~L. Everett, P.~Langacker and J.~Wang,
  \emph{{Physics implications of flat directions in free fermionic superstring
  models. 2. Renormalization group analysis}},
  \href{https://doi.org/10.1103/PhysRevD.59.115003}{\emph{Phys. Rev.}
  {\bfseries D59} (1999) 115003},
  [\href{https://arxiv.org/abs/hep-ph/9811355}{{\ttfamily hep-ph/9811355}}].

\bibitem{Anastasopoulos:2006da}
P.~Anastasopoulos, T.~P.~T. Dijkstra, E.~Kiritsis and A.~N. Schellekens,
  \emph{{Orientifolds, hypercharge embeddings and the Standard Model}},
  \href{https://doi.org/10.1016/j.nuclphysb.2006.10.013}{\emph{Nucl. Phys.}
  {\bfseries B759} (2006) 83--146},
  [\href{https://arxiv.org/abs/hep-th/0605226}{{\ttfamily hep-th/0605226}}].

\bibitem{Cvetic:2011iq}
M.~Cvetic, J.~Halverson and P.~Langacker, \emph{{Implications of String
  Constraints for Exotic Matter and Z' s Beyond the Standard Model}},
  \href{https://doi.org/10.1007/JHEP11(2011)058}{\emph{JHEP} {\bfseries 11}
  (2011) 058}, [\href{https://arxiv.org/abs/1108.5187}{{\ttfamily 1108.5187}}].

\bibitem{Suematsu:1994qm}
D.~Suematsu and Y.~Yamagishi, \emph{{Radiative symmetry breaking in a
  supersymmetric model with an extra U(1)}},
  \href{https://doi.org/10.1142/S0217751X95002096}{\emph{Int. J. Mod. Phys.}
  {\bfseries A10} (1995) 4521--4536},
  [\href{https://arxiv.org/abs/hep-ph/9411239}{{\ttfamily hep-ph/9411239}}].

\bibitem{Cvetic:1995rj}
M.~Cvetic and P.~Langacker, \emph{{Implications of Abelian extended gauge
  structures from string models}},
  \href{https://doi.org/10.1103/PhysRevD.54.3570}{\emph{Phys. Rev.} {\bfseries
  D54} (1996) 3570--3579},
  [\href{https://arxiv.org/abs/hep-ph/9511378}{{\ttfamily hep-ph/9511378}}].

\bibitem{King:2005jy}
S.~F. King, S.~Moretti and R.~Nevzorov, \emph{{Theory and phenomenology of an
  exceptional supersymmetric standard model}},
  \href{https://doi.org/10.1103/PhysRevD.73.035009}{\emph{Phys. Rev.}
  {\bfseries D73} (2006) 035009},
  [\href{https://arxiv.org/abs/hep-ph/0510419}{{\ttfamily hep-ph/0510419}}].

\bibitem{Sakharov:1967dj}
A.~D. Sakharov, \emph{{Violation of $CP$ invariance, $C$ asymmetry, and baryon
  asymmetry of the universe}},
  \href{https://doi.org/10.1070/PU1991v034n05ABEH002497}{\emph{Pis'ma Zh. Eksp.
  Teor. Fiz.} {\bfseries 5} (1967) 32--35}.

\bibitem{Cohen:1993nk}
A.~G. Cohen, D.~B. Kaplan and A.~E. Nelson, \emph{{Progress in electroweak
  baryogenesis}},
  \href{https://doi.org/10.1146/annurev.ns.43.120193.000331}{\emph{Ann. Rev.
  Nucl. Part. Sci.} {\bfseries 43} (1993) 27--70},
  [\href{https://arxiv.org/abs/hep-ph/9302210}{{\ttfamily hep-ph/9302210}}].

\bibitem{Trodden:1998ym}
M.~Trodden, \emph{{Electroweak baryogenesis}},
  \href{https://doi.org/10.1103/RevModPhys.71.1463}{\emph{Rev. Mod. Phys.}
  {\bfseries 71} (1999) 1463--1500},
  [\href{https://arxiv.org/abs/hep-ph/9803479}{{\ttfamily hep-ph/9803479}}].

\bibitem{Morrissey:2012db}
D.~E. Morrissey and M.~J. Ramsey-Musolf, \emph{{Electroweak baryogenesis}},
  \href{https://doi.org/10.1088/1367-2630/14/12/125003}{\emph{New J. Phys.}
  {\bfseries 14} (2012) 125003},
  [\href{https://arxiv.org/abs/1206.2942}{{\ttfamily 1206.2942}}].

\bibitem{White:2016nbo}
G.~A. White, \emph{{A Pedagogical Introduction to Electroweak Baryogenesis}}.
\newblock IOP Concise Physics. Morgan \& Claypool, New York, 2016,
  \href{https://doi.org/10.1088/978-1-6817-4457-5}{10.1088/978-1-6817-4457-5}.

\bibitem{Balazs:2007pf}
C.~Balazs, M.~Carena, A.~Freitas and C.~E.~M. Wagner, \emph{{Phenomenology of
  the nMSSM from colliders to cosmology}},
  \href{https://doi.org/10.1088/1126-6708/2007/06/066}{\emph{JHEP} {\bfseries
  06} (2007) 066}, [\href{https://arxiv.org/abs/0705.0431}{{\ttfamily
  0705.0431}}].

\bibitem{Balazs:2004ae}
C.~Bal\'azs, M.~Carena, A.~Menon, D.~E. Morrissey and C.~E.~M. Wagner,
  \emph{{Supersymmetric origin of matter}},
  \href{https://doi.org/10.1103/PhysRevD.71.075002}{\emph{Phys. Rev. D}
  {\bfseries 71} (2005) 075002},
  [\href{https://arxiv.org/abs/hep-ph/0412264}{{\ttfamily hep-ph/0412264}}].

\bibitem{Balazs:2013cia}
{\relax Cs}.~Bal\'azs, A.~Mazumdar, E.~Pukartas and G.~White,
  \emph{{Baryogenesis, dark matter and inflation in the next-to-minimal
  supersymmetric standard model}},
  \href{https://doi.org/10.1007/JHEP01(2014)073}{\emph{JHEP} {\bfseries 01}
  (2014) 073}, [\href{https://arxiv.org/abs/1309.5091}{{\ttfamily 1309.5091}}].

\bibitem{Athron:2019teq}
P.~Athron, C.~Balazs, A.~Fowlie, G.~Pozzo, G.~White and Y.~Zhang, \emph{{Strong
  first-order phase transitions in the NMSSM — a comprehensive survey}},
  \href{https://doi.org/10.1007/JHEP11(2019)151}{\emph{JHEP} {\bfseries 11}
  (2019) 151}, [\href{https://arxiv.org/abs/1908.11847}{{\ttfamily
  1908.11847}}].

\bibitem{Maggiore:1999vm}
M.~Maggiore, \emph{{Gravitational wave experiments and early universe
  cosmology}}, \href{https://doi.org/10.1016/S0370-1573(99)00102-7}{\emph{Phys.
  Rept.} {\bfseries 331} (2000) 283--367},
  [\href{https://arxiv.org/abs/gr-qc/9909001}{{\ttfamily gr-qc/9909001}}].

\bibitem{Weir:2017wfa}
D.~J. Weir, \emph{{Gravitational waves from a first order electroweak phase
  transition: a brief review}},
  \href{https://doi.org/10.1098/rsta.2017.0126}{\emph{Phil. Trans. Roy. Soc.
  Lond.} {\bfseries A376} (2018) 20170126},
  [\href{https://arxiv.org/abs/1705.01783}{{\ttfamily 1705.01783}}].

\bibitem{Alanne:2019bsm}
T.~Alanne, T.~Hugle, M.~Platscher and K.~Schmitz, \emph{{A fresh look at the
  gravitational-wave signal from cosmological phase transitions}},
  \href{https://doi.org/10.1007/JHEP03(2020)004}{\emph{JHEP} {\bfseries 03}
  (2020) 004}, [\href{https://arxiv.org/abs/1909.11356}{{\ttfamily
  1909.11356}}].

\bibitem{Schmitz:2020syl}
K.~Schmitz, \emph{{New Sensitivity Curves for Gravitational-Wave Experiments}},
   \href{https://arxiv.org/abs/2002.04615}{{\ttfamily 2002.04615}}.

\bibitem{Abbott:2016blz}
{\scshape LIGO Scientific, Virgo} collaboration, B.~P. Abbott et~al.,
  \emph{{Observation of Gravitational Waves from a Binary Black Hole Merger}},
  \href{https://doi.org/10.1103/PhysRevLett.116.061102}{\emph{Phys. Rev. Lett.}
  {\bfseries 116} (2016) 061102},
  [\href{https://arxiv.org/abs/1602.03837}{{\ttfamily 1602.03837}}].

\bibitem{TheLIGOScientific:2017qsa}
{\scshape LIGO Scientific, Virgo} collaboration, B.~P. Abbott et~al.,
  \emph{{GW170817: Observation of Gravitational Waves from a Binary Neutron
  Star Inspiral}},
  \href{https://doi.org/10.1103/PhysRevLett.119.161101}{\emph{Phys. Rev. Lett.}
  {\bfseries 119} (2017) 161101},
  [\href{https://arxiv.org/abs/1710.05832}{{\ttfamily 1710.05832}}].

\bibitem{Abbott:2016nmj}
{\scshape LIGO Scientific, Virgo} collaboration, B.~P. Abbott et~al.,
  \emph{{GW151226: Observation of Gravitational Waves from a 22-Solar-Mass
  Binary Black Hole Coalescence}},
  \href{https://doi.org/10.1103/PhysRevLett.116.241103}{\emph{Phys. Rev. Lett.}
  {\bfseries 116} (2016) 241103},
  [\href{https://arxiv.org/abs/1606.04855}{{\ttfamily 1606.04855}}].

\bibitem{Abbott:2017vtc}
{\scshape LIGO Scientific, VIRGO} collaboration, B.~P. Abbott et~al.,
  \emph{{GW170104: Observation of a 50-Solar-Mass Binary Black Hole Coalescence
  at Redshift 0.2}}, \href{https://doi.org/10.1103/PhysRevLett.118.221101,
  10.1103/PhysRevLett.121.129901}{\emph{Phys. Rev. Lett.} {\bfseries 118}
  (2017) 221101}, [\href{https://arxiv.org/abs/1706.01812}{{\ttfamily
  1706.01812}}].

\bibitem{Abbott:2017oio}
{\scshape LIGO Scientific, Virgo} collaboration, B.~P. Abbott et~al.,
  \emph{{GW170814: A Three-Detector Observation of Gravitational Waves from a
  Binary Black Hole Coalescence}},
  \href{https://doi.org/10.1103/PhysRevLett.119.141101}{\emph{Phys. Rev. Lett.}
  {\bfseries 119} (2017) 141101},
  [\href{https://arxiv.org/abs/1709.09660}{{\ttfamily 1709.09660}}].

\bibitem{Vaskonen:2016yiu}
V.~Vaskonen, \emph{{Electroweak baryogenesis and gravitational waves from a
  real scalar singlet}},
  \href{https://doi.org/10.1103/PhysRevD.95.123515}{\emph{Phys. Rev.}
  {\bfseries D95} (2017) 123515},
  [\href{https://arxiv.org/abs/1611.02073}{{\ttfamily 1611.02073}}].

\bibitem{Alves:2018jsw}
A.~Alves, T.~Ghosh, H.-K. Guo, K.~Sinha and D.~Vagie, \emph{{Collider and
  Gravitational Wave Complementarity in Exploring the Singlet Extension of the
  Standard Model}}, \href{https://doi.org/10.1007/JHEP04(2019)052}{\emph{JHEP}
  {\bfseries 04} (2019) 052},
  [\href{https://arxiv.org/abs/1812.09333}{{\ttfamily 1812.09333}}].

\bibitem{Alves:2019igs}
A.~Alves, D.~Gonçalves, T.~Ghosh, H.-K. Guo and K.~Sinha, \emph{{Di-Higgs
  Production in the $4b$ Channel and Gravitational Wave Complementarity}},
  \href{https://arxiv.org/abs/1909.05268}{{\ttfamily 1909.05268}}.

\bibitem{Hashino:2016xoj}
K.~Hashino, M.~Kakizaki, S.~Kanemura, P.~Ko and T.~Matsui, \emph{{Gravitational
  waves and Higgs boson couplings for exploring first order phase transition in
  the model with a singlet scalar field}},
  \href{https://doi.org/10.1016/j.physletb.2016.12.052}{\emph{Phys. Lett.}
  {\bfseries B766} (2017) 49--54},
  [\href{https://arxiv.org/abs/1609.00297}{{\ttfamily 1609.00297}}].

\bibitem{Beniwal:2017eik}
A.~Beniwal, M.~Lewicki, J.~D. Wells, M.~White and A.~G. Williams,
  \emph{{Gravitational wave, collider and dark matter signals from a scalar
  singlet electroweak baryogenesis}},
  \href{https://doi.org/10.1007/JHEP08(2017)108}{\emph{JHEP} {\bfseries 08}
  (2017) 108}, [\href{https://arxiv.org/abs/1702.06124}{{\ttfamily
  1702.06124}}].

\bibitem{Kang:2017mkl}
Z.~Kang, P.~Ko and T.~Matsui, \emph{{Strong first order EWPT \& strong
  gravitational waves in Z$_{3}$-symmetric singlet scalar extension}},
  \href{https://doi.org/10.1007/JHEP02(2018)115}{\emph{JHEP} {\bfseries 02}
  (2018) 115}, [\href{https://arxiv.org/abs/1706.09721}{{\ttfamily
  1706.09721}}].

\bibitem{Beniwal:2018hyi}
A.~Beniwal, M.~Lewicki, M.~White and A.~G. Williams, \emph{{Gravitational waves
  and electroweak baryogenesis in a global study of the extended scalar singlet
  model}}, \href{https://doi.org/10.1007/JHEP02(2019)183}{\emph{JHEP}
  {\bfseries 02} (2019) 183},
  [\href{https://arxiv.org/abs/1810.02380}{{\ttfamily 1810.02380}}].

\bibitem{Croon:2019kpe}
D.~Croon, T.~E. Gonzalo, L.~Graf, N.~Košnik and G.~White, \emph{{GUT Physics
  in the era of the LHC}},
  \href{https://doi.org/10.3389/fphy.2019.00076}{\emph{Front.in Phys.}
  {\bfseries 7} (2019) 76}, [\href{https://arxiv.org/abs/1903.04977}{{\ttfamily
  1903.04977}}].

\bibitem{Dev:2016feu}
P.~S.~B. Dev and A.~Mazumdar, \emph{{Probing the Scale of New Physics by
  Advanced LIGO/VIRGO}},
  \href{https://doi.org/10.1103/PhysRevD.93.104001}{\emph{Phys. Rev.}
  {\bfseries D93} (2016) 104001},
  [\href{https://arxiv.org/abs/1602.04203}{{\ttfamily 1602.04203}}].

\bibitem{Dev:2019njv}
P.~S.~B. Dev, F.~Ferrer, Y.~Zhang and Y.~Zhang, \emph{{Gravitational Waves from
  First-Order Phase Transition in a Simple Axion-Like Particle Model}},
  \href{https://doi.org/10.1088/1475-7516/2019/11/006}{\emph{JCAP} {\bfseries
  1911} (2019) 006}, [\href{https://arxiv.org/abs/1905.00891}{{\ttfamily
  1905.00891}}].

\bibitem{Bian:2019zpn}
L.~Bian, H.-K. Guo, Y.~Wu and R.~Zhou, \emph{{Gravitational wave and collider
  searches for electroweak symmetry breaking patterns}},
  \href{https://doi.org/10.1103/PhysRevD.101.035011}{\emph{Phys. Rev.}
  {\bfseries D101} (2020) 035011},
  [\href{https://arxiv.org/abs/1906.11664}{{\ttfamily 1906.11664}}].

\bibitem{Bian:2019kmg}
L.~Bian, Y.~Wu and K.-P. Xie, \emph{{Electroweak phase transition with
  composite Higgs models: calculability, gravitational waves and collider
  searches}}, \href{https://doi.org/10.1007/JHEP12(2019)028}{\emph{JHEP}
  {\bfseries 12} (2019) 028},
  [\href{https://arxiv.org/abs/1909.02014}{{\ttfamily 1909.02014}}].

\bibitem{Punturo:2010zz}
M.~Punturo et~al., \emph{{The Einstein Telescope: A third-generation
  gravitational wave observatory}},
  \href{https://doi.org/10.1088/0264-9381/27/19/194002}{\emph{Class. Quant.
  Grav.} {\bfseries 27} (2010) 194002}.

\bibitem{TheLIGOScientific:2014jea}
{\scshape LIGO Scientific} collaboration, J.~Aasi et~al., \emph{{Advanced
  LIGO}}, \href{https://doi.org/10.1088/0264-9381/32/7/074001}{\emph{Class.
  Quant. Grav.} {\bfseries 32} (2015) 074001},
  [\href{https://arxiv.org/abs/1411.4547}{{\ttfamily 1411.4547}}].

\bibitem{TheVirgo:2014hva}
{\scshape VIRGO} collaboration, F.~Acernese et~al., \emph{{Advanced Virgo: a
  second-generation interferometric gravitational wave detector}},
  \href{https://doi.org/10.1088/0264-9381/32/2/024001}{\emph{Class. Quant.
  Grav.} {\bfseries 32} (2015) 024001},
  [\href{https://arxiv.org/abs/1408.3978}{{\ttfamily 1408.3978}}].

\bibitem{Akutsu:2018axf}
{\scshape KAGRA} collaboration, T.~Akutsu et~al., \emph{{KAGRA: 2.5 Generation
  Interferometric Gravitational Wave Detector}},
  \href{https://doi.org/10.1038/s41550-018-0658-y}{\emph{Nat. Astron.}
  {\bfseries 3} (2019) 35--40},
  [\href{https://arxiv.org/abs/1811.08079}{{\ttfamily 1811.08079}}].

\bibitem{Reitze:2019iox}
D.~Reitze et~al., \emph{{Cosmic Explorer: The U.S. Contribution to
  Gravitational-Wave Astronomy beyond LIGO}}, {\emph{Bull. Am. Astron. Soc.}
  {\bfseries 51} (2019) 035},
  [\href{https://arxiv.org/abs/1907.04833}{{\ttfamily 1907.04833}}].

\bibitem{amaroseoane2017laser}
P.~Amaro-Seoane, H.~Audley, S.~Babak, J.~Baker, E.~Barausse, P.~Bender et~al.,
  \emph{Laser interferometer space antenna},
  \href{https://arxiv.org/abs/1702.00786}{{\ttfamily 1702.00786}}.

\bibitem{Kawamura:2006up}
S.~Kawamura et~al., \emph{{The Japanese space gravitational wave antenna
  DECIGO}}, \href{https://doi.org/10.1088/0264-9381/23/8/S17}{\emph{Class.
  Quant. Grav.} {\bfseries 23} (2006) S125--S132}.

\bibitem{Harry:2006fi}
G.~M. Harry, P.~Fritschel, D.~A. Shaddock, W.~Folkner and E.~S. Phinney,
  \emph{{Laser interferometry for the big bang observer}},
  \href{https://doi.org/10.1088/0264-9381/23/24/C01,
  10.1088/0264-9381/23/15/008}{\emph{Class. Quant. Grav.} {\bfseries 23} (2006)
  4887--4894}.

\bibitem{Hu:2017mde}
W.-R. Hu and Y.-L. Wu, \emph{{The Taiji Program in Space for gravitational wave
  physics and the nature of gravity}},
  \href{https://doi.org/10.1093/nsr/nwx116}{\emph{Natl. Sci. Rev.} {\bfseries
  4} (2017) 685--686}.

\bibitem{Croon:2018kqn}
D.~Croon, T.~E. Gonzalo and G.~White, \emph{{Gravitational Waves from a
  Pati-Salam Phase Transition}},
  \href{https://doi.org/10.1007/JHEP02(2019)083}{\emph{JHEP} {\bfseries 02}
  (2019) 083}, [\href{https://arxiv.org/abs/1812.02747}{{\ttfamily
  1812.02747}}].

\bibitem{Apreda:2001us}
R.~Apreda, M.~Maggiore, A.~Nicolis and A.~Riotto, \emph{{Gravitational waves
  from electroweak phase transitions}},
  \href{https://doi.org/10.1016/S0550-3213(02)00264-X}{\emph{Nucl. Phys.}
  {\bfseries B631} (2002) 342--368},
  [\href{https://arxiv.org/abs/gr-qc/0107033}{{\ttfamily gr-qc/0107033}}].

\bibitem{Leitao:2012tx}
L.~Leitao, A.~Megevand and A.~D. Sanchez, \emph{{Gravitational waves from the
  electroweak phase transition}},
  \href{https://doi.org/10.1088/1475-7516/2012/10/024}{\emph{JCAP} {\bfseries
  1210} (2012) 024}, [\href{https://arxiv.org/abs/1205.3070}{{\ttfamily
  1205.3070}}].

\bibitem{Dorsch:2018pat}
G.~C. Dorsch, S.~J. Huber and T.~Konstandin, \emph{{Bubble wall velocities in
  the Standard Model and beyond}},
  \href{https://doi.org/10.1088/1475-7516/2018/12/034}{\emph{JCAP} {\bfseries
  1812} (2018) 034}, [\href{https://arxiv.org/abs/1809.04907}{{\ttfamily
  1809.04907}}].

\bibitem{Ellis:2018mja}
J.~Ellis, M.~Lewicki and J.~M. No, \emph{{On the Maximal Strength of a
  First-Order Electroweak Phase Transition and its Gravitational Wave Signal}},
   \href{https://arxiv.org/abs/1809.08242}{{\ttfamily 1809.08242}}.

\bibitem{Ellis:2019oqb}
J.~Ellis, M.~Lewicki, J.~M. No and V.~Vaskonen, \emph{{Gravitational wave
  energy budget in strongly supercooled phase transitions}},
  \href{https://doi.org/10.1088/1475-7516/2019/06/024}{\emph{JCAP} {\bfseries
  1906} (2019) 024}, [\href{https://arxiv.org/abs/1903.09642}{{\ttfamily
  1903.09642}}].

\bibitem{Wainwright:CosmoTransition}
C.~L. Wainwright, \emph{{CosmoTransitions: Computing Cosmological Phase
  Transition Temperatures and Bubble Profiles with Multiple Fields}},
  \href{https://doi.org/10.1016/j.cpc.2012.04.004}{\emph{Comput. Phys. Commun.}
  {\bfseries 183} (2012) 2006--2013},
  [\href{https://arxiv.org/abs/1109.4189}{{\ttfamily 1109.4189}}].

\bibitem{Basler:2018cwe}
P.~Basler and M.~Mühlleitner, \emph{{BSMPT (Beyond the Standard Model Phase
  Transitions): A tool for the electroweak phase transition in extended Higgs
  sectors}}, \href{https://doi.org/10.1016/j.cpc.2018.11.006}{\emph{Comput.
  Phys. Commun.} {\bfseries 237} (2019) 62--85},
  [\href{https://arxiv.org/abs/1803.02846}{{\ttfamily 1803.02846}}].

\bibitem{Camargo-Molina:2013qva}
J.~E. Camargo-Molina, B.~O'Leary, W.~Porod and F.~Staub,
  \emph{{$\mathbf{Vevacious}$: A Tool For Finding The Global Minima Of One-Loop
  Effective Potentials With Many Scalars}},
  \href{https://doi.org/10.1140/epjc/s10052-013-2588-2}{\emph{Eur. Phys. J.}
  {\bfseries C73} (2013) 2588},
  [\href{https://arxiv.org/abs/1307.1477}{{\ttfamily 1307.1477}}].

\bibitem{Athron:2014yba}
P.~Athron, J.-h. Park, D.~Stöckinger and A.~Voigt, \emph{{FlexibleSUSY -- A
  spectrum generator generator for supersymmetric models}},
  \href{https://doi.org/10.1016/j.cpc.2014.12.020}{\emph{Comput. Phys. Commun.}
  {\bfseries 190} (2015) 139--172},
  [\href{https://arxiv.org/abs/1406.2319}{{\ttfamily 1406.2319}}].

\bibitem{Athron:2016fuq}
P.~Athron, J.-h. Park, T.~Steudtner, D.~Stöckinger and A.~Voigt,
  \emph{{Precise Higgs mass calculations in (non-)minimal supersymmetry at both
  high and low scales}},
  \href{https://doi.org/10.1007/JHEP01(2017)079}{\emph{JHEP} {\bfseries 01}
  (2017) 079}, [\href{https://arxiv.org/abs/1609.00371}{{\ttfamily
  1609.00371}}].

\bibitem{Athron:2017fvs}
P.~Athron, M.~Bach, D.~Harries, T.~Kwasnitza, J.-h. Park, D.~Stöckinger
  et~al., \emph{{FlexibleSUSY 2.0: Extensions to investigate the phenomenology
  of SUSY and non-SUSY models}},
  \href{https://doi.org/10.1016/j.cpc.2018.04.016}{\emph{Comput. Phys. Commun.}
  {\bfseries 230} (2018) 145--217},
  [\href{https://arxiv.org/abs/1710.03760}{{\ttfamily 1710.03760}}].

\bibitem{JAMES1975343}
F.~James and M.~Roos, \emph{Minuit - a system for function minimization and
  analysis of the parameter errors and correlations},
  \href{https://doi.org/https://doi.org/10.1016/0010-4655(75)90039-9}{\emph{Computer
  Physics Communications} {\bfseries 10} (1975) 343 -- 367}.

\bibitem{Patel:2011th}
H.~H. Patel and M.~J. Ramsey-Musolf, \emph{{Baryon Washout, Electroweak Phase
  Transition, and Perturbation Theory}},
  \href{https://doi.org/10.1007/JHEP07(2011)029}{\emph{JHEP} {\bfseries 07}
  (2011) 029}, [\href{https://arxiv.org/abs/1101.4665}{{\ttfamily 1101.4665}}].

\bibitem{Martin:2001vx}
S.~P. Martin, \emph{{Two Loop Effective Potential for a General Renormalizable
  Theory and Softly Broken Supersymmetry}},
  \href{https://doi.org/10.1103/PhysRevD.65.116003}{\emph{Phys. Rev.}
  {\bfseries D65} (2002) 116003},
  [\href{https://arxiv.org/abs/hep-ph/0111209}{{\ttfamily hep-ph/0111209}}].

\bibitem{Siegel:1979wq}
W.~Siegel, \emph{{Supersymmetric Dimensional Regularization via Dimensional
  Reduction}}, \href{https://doi.org/10.1016/0370-2693(79)90282-X}{\emph{Phys.
  Lett.} {\bfseries 84B} (1979) 193--196}.

\bibitem{Capper:1979ns}
D.~M. Capper, D.~R.~T. Jones and P.~van Nieuwenhuizen, \emph{{Regularization by
  Dimensional Reduction of Supersymmetric and Nonsupersymmetric Gauge
  Theories}}, \href{https://doi.org/10.1016/0550-3213(80)90244-8}{\emph{Nucl.
  Phys.} {\bfseries B167} (1980) 479--499}.

\bibitem{Jack:1994rk}
I.~Jack, D.~R.~T. Jones, S.~P. Martin, M.~T. Vaughn and Y.~Yamada,
  \emph{{Decoupling of the epsilon scalar mass in softly broken
  supersymmetry}}, \href{https://doi.org/10.1103/PhysRevD.50.R5481}{\emph{Phys.
  Rev.} {\bfseries D50} (1994) R5481--R5483},
  [\href{https://arxiv.org/abs/hep-ph/9407291}{{\ttfamily hep-ph/9407291}}].

\bibitem{Garny:2012cg}
M.~Garny and T.~Konstandin, \emph{{On the gauge dependence of vacuum
  transitions at finite temperature}},
  \href{https://doi.org/10.1007/JHEP07(2012)189}{\emph{JHEP} {\bfseries 07}
  (2012) 189}, [\href{https://arxiv.org/abs/1205.3392}{{\ttfamily 1205.3392}}].

\bibitem{Linde:1980ts}
A.~D. Linde, \emph{{Infrared Problem in Thermodynamics of the Yang-Mills Gas}},
  \href{https://doi.org/10.1016/0370-2693(80)90769-8}{\emph{Phys. Lett.}
  {\bfseries 96B} (1980) 289--292}.

\bibitem{Gross:1980br}
D.~J. Gross, R.~D. Pisarski and L.~G. Yaffe, \emph{{QCD and Instantons at
  Finite Temperature}},
  \href{https://doi.org/10.1103/RevModPhys.53.43}{\emph{Rev. Mod. Phys.}
  {\bfseries 53} (1981) 43}.

\bibitem{Coleman:1977py}
S.~R. Coleman, \emph{{The Fate of the False Vacuum. 1. Semiclassical Theory}},
  \href{https://doi.org/10.1103/PhysRevD.15.2929,
  10.1103/PhysRevD.16.1248}{\emph{Phys. Rev.} {\bfseries D15} (1977)
  2929--2936}.

\bibitem{Callan:1977pt}
C.~G. Callan, Jr. and S.~R. Coleman, \emph{{The Fate of the False Vacuum. 2.
  First Quantum Corrections}},
  \href{https://doi.org/10.1103/PhysRevD.16.1762}{\emph{Phys. Rev.} {\bfseries
  D16} (1977) 1762--1768}.

\bibitem{Linde:1980tt}
A.~D. Linde, \emph{{Fate of the False Vacuum at Finite Temperature: Theory and
  Applications}},
  \href{https://doi.org/10.1016/0370-2693(81)90281-1}{\emph{Phys. Lett.}
  {\bfseries 100B} (1981) 37--40}.

\bibitem{Athron:2019nbd}
P.~Athron, C.~Balázs, M.~Bardsley, A.~Fowlie, D.~Harries and G.~White,
  \emph{{BubbleProfiler: finding the field profile and action for cosmological
  phase transitions}},
  \href{https://doi.org/10.1016/j.cpc.2019.05.017}{\emph{Comput. Phys. Commun.}
  {\bfseries 244} (2019) 448--468},
  [\href{https://arxiv.org/abs/1901.03714}{{\ttfamily 1901.03714}}].

\bibitem{Quiros:2007zz}
M.~Quir\'os, \emph{{Field Theory at Finite Temperature and Phase Transitions}},
  {\emph{Acta Phys. Polon. B} {\bfseries 38} (2007) 3661--3703}.

\bibitem{Caprini:2019egz}
C.~Caprini et~al., \emph{{Detecting gravitational waves from cosmological phase
  transitions with LISA: an update}},
  \href{https://doi.org/10.1088/1475-7516/2020/03/024}{\emph{JCAP} {\bfseries
  03} (2020) 024}, [\href{https://arxiv.org/abs/1910.13125}{{\ttfamily
  1910.13125}}].

\bibitem{Rowan90functionalstability}
T.~H. Rowan, \emph{Functional stability analysis of numerical algorithms},
  tech. rep., University of Texas, 1990.

\bibitem{nlopt}
S.~G. Johnson, ``{The \code{NLopt} nonlinear-optimization package}.''

\bibitem{Arnold:1992rz}
P.~B. Arnold and O.~Espinosa, \emph{{The Effective potential and first order
  phase transitions: Beyond leading-order}},
  \href{https://doi.org/10.1103/physrevd.50.6662.2,
  10.1103/PhysRevD.47.3546}{\emph{Phys. Rev.} {\bfseries D47} (1993) 3546},
  [\href{https://arxiv.org/abs/hep-ph/9212235}{{\ttfamily hep-ph/9212235}}].

\end{thebibliography}\endgroup

\end{document}